\documentclass[]{aa} % for the letters
\usepackage{graphicx}
\usepackage{txfonts}
\begin{document}
%
% Personal definitions
\def\hi {H\,{\sc i}}
\def\hii {H\,{\sc ii}}
\def\hdueo {H$_2$O}
\def\meth {CH$_{3}$OH}
\def\dg{$^{\circ}$}
\def\kms{km\,s$^{-1}$}
\def\jyb{Jy\,beam$^{-1}$}
\def\mjyb{mJy\,beam$^{-1}$}
\def\solmass {\hbox{M$_{\odot}$}}
\def\solum {\hbox{L$_{\odot}$}} 
\def\d {$^{\circ}$}
\def\n {$n_{\rm{H_{2}}}$}
\title{The structure of the magnetic field in the massive star-forming region W75N}

\author{G.\ Surcis \thanks{Member of the International Max Planck Research School (IMPRS) for Astronomy and Astrophysics at the Universities of Bonn and Cologne.}  \inst{, 1}
  \and 
   W.H.T. \ Vlemmings \inst{1}
  \and
  S. Curiel \inst{2}
  \and
  B. Hutawarakorn Kramer \inst{3,4}
  \and
  J.M. Torrelles \inst{5}
  \and
  A.P. Sarma \inst{6}
  }

\institute{ Argelander-Institut f\"{u}r Astronomie der Universit\"{a}t Bonn, Auf dem H\"{u}gel 71, D-53121 Bonn, Germany\\
 \email{gsurcis@astro.uni-bonn.de}
 \and
 Instituto de Astronom\'{\i}a (UNA), Apdo Postal 70-264, Cd. Universitaria, 04510-Mexico DF, Mexico 
 \and
 Max-Planck Institut f\"{u}r Radioastronomie, Auf dem H\"{u}gel 69, D-53121 Bonn, Germany
 \and
 National Astronomical Research Institute of Thailand, Ministry of Science and Technology, Rama VI Rd., Bangkok 10400, Thailand
 \and
  Instituto de Ciencias del Espacio (CSIC)-UB/IEEC, Universitat de Barcelona, Mart\'{\i} i Franqu\`{e}s 1, E-08028 Barcelona, Spain
 \and
  Physics Department, DePaul University, 2219 N. Kenmore Ave., Byrne Hall 211, Chicago, IL 60614, USA 
  }

\date{Received ; accepted}
\abstract
%context heading (optional)
{A debated topic in star formation theory is the role of magnetic fields during the protostellar phase of high-mass stars. 
It is still unclear how magnetic fields influence the formation and dynamics of massive disks and outflows. Most current information 
on magnetic fields close to high-mass protostars comes from polarized maser emissions, which allows us to investigate the magnetic field 
on small scales by using very long-baseline interferometry.}
% aims heading (mandatory)
{The massive star-forming region W75N contains three radio continuum sources (VLA\,1, VLA\,2, and VLA\,3), at three different evolutionary
 stages, and associated masers, while a large-scale molecular bipolar outflow is also present. Very recently, polarization observations of
 the 6.7\,GHz methanol masers at milliarsecond resolution have been able to probe the strength and structure of the magnetic field over more than 2000 AU 
around VLA\,1. The magnetic field is parallel to the outflow, suggesting that VLA\,1 is its powering source. The observations of \hdueo \, 
masers at 22\,GHz can give more information about the gas dynamics and the magnetic fields around VLA\,1 and VLA\,2.} 
% methods heading (mandatory)
{The NRAO Very Long Baseline Array was used to measure the linear polarization and the Zeeman-splitting of the 22\,GHz water masers in the
 star-forming region W75N. }
% results heading (mandatory)
{We detected 124 water masers, 36 around VLA\,1 and 88 around VLA\,2 of W75N, which indicate two different physical environments around 
the two sources, where VLA\,1 is in a more evolved state. The linear polarization of the masers confirms the tightly ordered magnetic field 
around VLA\,1, which is aligned with the large-scale molecular outflow, and also reveals an ordered magnetic field around VLA\,2, which 
is not parallel to the outflow. The Zeeman-splitting measured on 20 of the masers indicates strong magnetic fields around both sources 
(the averaged values are $|B_{\rm{VLA1}}|\sim700$\,mG and $|B_{\rm{VLA2}}|\sim1700$\,mG). 
The high values of the magnetic field strengths, which come from the shock compression of the gas,
 are consistent with the methanol and OH magnetic field strengths. 
Moreover, by studying the maser properties we were also able to determine that the water masers are pumped in
C-shocks in both sources.}
{}

\titlerunning{W75N: water masers polarization.}
\authorrunning{Surcis et al.}

\maketitle
%________________________________________________________________
\section{Introduction}
During the formation of low-mass stars, the magnetic field is thought to slow the collapse, to transfer the angular momentum, and to power 
the outflow, but its role during the formation of high-mass stars is still under debate (e.g., McKee \& Ostriker \cite{McK07}, Girart et 
al. \cite{gir09}). Several questions remain unanswered because massive star-forming regions are rare and distant and often consist of a large
 number of protostars, consequently it is difficult to observe magnetic fields during their protostellar phase. Recently some progress has
 been made. Vlemmings et al. (\cite{vle10}) and Surcis et al. (\cite{sur09}, hereafter S09) have shown that the magnetic field is orthogonal 
to large rotating disks and parallel to molecular bipolar outflows even during the high-mass protostellar phase and not only during the 
formation of low-mass stars (e.g., Matsumoto \& Tomisaka \cite{mat04}, McKee \& Ostriker \cite{McK07}). They obtained their results by 
using methanol masers as probes of magnetic fields. In fact, the bright and narrow spectral line emission of masers is ideal for 
detecting, with polarimetric interferometry, both Zeeman-splitting and the orientation of the magnetic field on scales from arcsec to 
milliarcsec (mas). So far, the main maser species used for this purpose are \hdueo, OH, and \meth \, masers. Since their masing conditions 
are different they are located in distinct zones of massive star-forming regions. The \hdueo \, masers are detected in the denser zones
 with hydrogen number densities \n \, between approximately 10$^{8}$ and 10$^{10}$ cm$^{-3}$ (Elitzur et al. \cite{eli89}), showing a 
velocity width of about 1 \kms \, and a high brightness temperature $T_{\rm{b}}>10^{9} \rm{K}$ (e.g., Reid \& Moran \cite{rei81}). The 
first very long baseline interferometry (VLBI) observations of the linear and circular polarization of water maser emissions were made 
by Lepp\"{a}nen et al. (\cite{lep98}) and Sarma et al.  (\cite{sar01}), respectively. Afterwards more \hdueo \, maser polarization 
observations were carried out by various authors, which confirmed the importance of this kind of observation in studying the role of 
the magnetic field in massive star formation (e.g., Imai et al. \cite{ima03}, Vlemmings et al. \cite{vle06}). Here we present Very Long 
Baseline Array (VLBA) linear and circular polarization observations of water masers in the high-mass star-forming region W75N.\\
\indent W75N is an active high-mass star-forming region in the molecular complex DR21--W75 (Dickel et al. \cite{dic78}, Persi et al. 
\cite{per06}) at a distance of 2\,kpc. At the resolution of $\sim1''\!\!\!.5$ three \hii \, regions were identified (Haschick et al. 
\cite{has81}): W75N\,(A), W75N\,(B), and W75N\,(C). At $0\farcs5$ resolution, Hunter et al. (\cite{hun94}) detected three subregions in 
W75N\,(B): Ba, Bb and Bc. At higher angular resolution ($\sim0\farcs1$), Torrelles et al. (\cite{tor97}) resolved the subregions Ba and 
Bb further (renaming them VLA\,1 and VLA\,3), and imaged a third weaker and more compact \hii \, region between them, which they named 
VLA\,2. A large-scale high-velocity outflow, with an extension greater than 3 pc and a total molecular mass greater than 255~\solmass,  
was also detected from W75N\,(B) (e.g., Shepherd et al. \cite{she03}).  Shepherd et al. (\cite{she03}) proposed a multi-outflow scenario 
where VLA\,2 may drive the large-scale outflow, and VLA\,1 and VLA\,3 are the centers of two other small flows. So far, it has been 
impossible to determine the main powering source of the 3 pc outflow. Several authors have suggested VLA\,1 as the powering source 
(e.g., Torrelles et al. \cite{tor97}, S09 and references therein).\\
\indent Several maser species (\hdueo \,, OH, and \meth) have been detected in W75N (e.g., Torrelles et al. \cite{tor97}, Baart et al 
\cite{baa86}, S09), in particular around the two H\,{\scriptsize II} regions VLA\,1 and VLA\,2. The \hdueo \, maser emissions associated 
with VLA\,1 are located along a linear structure parallel to its radio jet with a proper motion of about 2~mas/yr, while those associated 
with VLA\,2 show a circular distribution, which is expanding with a velocity of about 5 mas/yr (Torrelles et al., \cite{tor03}, hereafter 
T03). Only one \hdueo \, maser is associated with VLA\,3 (T03). Methanol masers were detected in two groups (A and B/C) located northwest 
and southeast of VLA\,1, respectively. Group A is along a linear structure and group B/C show an arc-like structure (S09). 
No \meth ~masers are associated with VLA\,2 (e.g., Minier et al. \cite{min00}, S09). However, VLA\,2 is the place where the most intensive 
OH flare took place (Alakov et al. \cite{ala05}, Slysh et al. \cite{sly10}).  Other OH maser emission sites are situated on a ring 
structure around all three radio sources (Hutawarakorn et al. \cite{hut02}). Based on the different activity in \hdueo \, and OH masers, 
Torrelles et al. (\cite{tor97}) suggest that these three sources are at different evolutionary stages, in particular VLA\,1 is the 
``oldest'' source of this region and VLA\,3 the ``youngest'' one (driving a radio Herbig-Haro object, Carrasco-Gonz\'{a}lez et al. 
\cite{car10}). In this evolutionary scenario, VLA\,2 is at an intermediate stage.\\
\indent In order to determine and investigate the magnetic field of W75N(B), polarization observations of OH and \meth \, maser emissions 
have been made (e.g., Hutawarakorn et al. \cite{hut02}; Fish \& Reid \cite{fis07}; S09). The magnetic field strength obtained from the OH 
maser polarization observations was  $\sim$7~mG (e.g., Hutawarakorn et al. \cite{hut02}; Slysh et al. \cite{sly02}).  During the OH maser 
flare near VLA\,2, Slysh \& Migenes (\cite{sly06}) detected a strong magnetic field of more than 40 mG in several maser spots, which 
increased in the next observations up to 70\,mG (Slysh et al. \cite{sly10}). S09 measured a magnetic field strength of 50 mG around VLA\,1
 by studying the circular polarized emissions of methanol masers. Hutawarakorn \& Cohen (\cite{hut96}) and S09 investigated the linear 
polarization of OH and \meth \, masers, respectively, and both suggested that the magnetic field is oriented along the outflow. Although 
the OH masers at 1.6 and 1.7\,GHz are very susceptible to both internal and external Faraday rotation, at 6.7 GHz the Faraday rotation is 
only about 5\d \,indicating that the finding of the large-scale magnetic field aligned with the outflow is robust.\\
\indent Since the water masers, unlike the \meth \, and OH masers, are found in denser zones of the star-forming regions, it is 
crucial to study their polarization emission on a small scale. In fact, because they arise close to both radio sources they can 
give us new information on the role of the magnetic field during the massive star formation, in particular at two different evolutionary 
stages. 
\section{Observations and data reduction}\label{obssect}
We observed the massive star-forming region W75N(B) in the 6$\rm{_{16}}$-5$\rm{_{23}}$ transition of \hdueo \, (rest frequency: 
22.23508~GHz) with the NRAO\footnote{The National Radio Astronomy Observatory (NRAO) is a facility of the National Science Foundation 
operated under cooperative agreement by Associated Universities, Inc.} VLBA on November 21$\rm{^{st}}$ 2005. The observations were made 
in full polarization spectral mode using 4 overlapped baseband filters of 1 MHz in order to cover a total velocity range of
$\approx50$~\kms. Two correlations were performed. One with 128 channels in order to generate all 4 polarization combinations (RR, LL, RL,
 LR) with a spectral resolution of 7.8~kHz (0.1~\kms). The other one with high spectral resolution (512 channels; 1.96~kHz=0.027~\kms), 
which only contains the circular polarization combinations (LL, RR), to be able to detect Zeeman splitting of the \hdueo \, maser across 
the entire velocity range. Including the overheads, the total observation time was 8~h.\\
\indent The data were reduced using the Astronomical Image Processing Software package (AIPS) following the method of Kemball et al. 
(\cite{kem95}). The bandpass, the delay, the phase and the polarization calibration were performed on the calibrator J2202+4216, which 
has been used successfully by Vlemmings et al. (\cite{vle06}) for \hdueo \, maser polarization observations of Cepheus\,A. The 
fringe-fitting and the self-calibration were performed on one of the brightest maser features VLA\,1.14 (Table~\ref{polt}). All 
calibration steps were initially performed on the dataset with modest spectral resolution after which the solutions, with the exception 
of the bandpass solutions that were obtained separately, were copied and applied to the high spectral resolution dataset. Stokes 
\textit{I} ($rms=7.3$~\mjyb), \textit{Q} ($rms=6.5$~\mjyb) and \textit{U} ($rms=6.5$~\mjyb) data cubes were created using the AIPS task 
IMAGR (beam-size 2.0~mas~$\times$~0.7 mas) from the modest spectral resolution dataset, while the \textit{I} and \textit{V} ($rms=10$~\mjyb) 
cubes were imaged from the high spectral resolution dataset and for the same fields. The \textit{Q} and \textit{U} cubes were combined
 to produce cubes of polarized intensity and polarization angle. Since these observations were obtained between two VLA polarization 
calibration observations\footnote{http://www.aoc.nrao.edu/astro/calib/polar}, during which the linear polarization angle of J2202+4216 was
 constant at 13.$\!\!$\dg5, we were able to estimate the polarization angles with a 
systemic error of no more than $\sim$2\dg. The \textit{I} and \textit{V} cubes at high spectral resolution were used to determine the 
Zeeman splitting. 
\section{Analysis}
\subsection{Maser identification}
To identify the water maser features in W75N(B) region, we used an identification process divided into three steps: 1) we used a 
program called ``maser finder'' (Curiel, personal communication), which is able to search for maser spots, velocity channel by velocity 
channel, with a signal-to-noise ratio greater than a given value (in our case, an 8-sigma limit was used); 2) we fitted the identified 
maser spots using the AIPS task IMFIT; and 3) we identified a maser feature, when three or more maser spots coincided spatially (within a 
box of 2 by 2 pixels) and each of them appeared in consecutive velocity channels. Table~\ref{polt} contains the positions, velocities and 
peak flux densities of all the maser features found in the region.\\
\indent In the first part of the process and after evaluating the global noise level of the channel map, the code searches for maser spots 
in the velocity map, excluding the edges of the channel map. When strong maser spots are found (for instance, with a peak flux density 
higher than 2\,\jyb), the local rms around these strong maser spots (within a predefined box, for instance, 400 by 400 pixels) is 
estimated, as well as the rms outside the boxes. The code identifies a maser spot when the SNR of the candidate (using the local rms) is 
greater than a predefined value. In the case of W75N, we have adopted a lower limit of 8-sigma. The code produces a table with the spatial 
location and intensity of all the identified maser spots. This procedure is repeated for each velocity channel included in the input file 
that is used by the code. The code also produces a script for each velocity channel that can be run inside AIPS in order to produce 
Gaussian fits of all the identified maser spots. After running all the AIPS scripts, a table with all the important parameters (e.g., 
spatial resolution, velocity, intensity) of each maser spot is produced. We then use the output table to search for maser features that 
fulfill the criteria described above.
\begin{figure*}[t!]
\centering
\includegraphics[width = 8 cm]{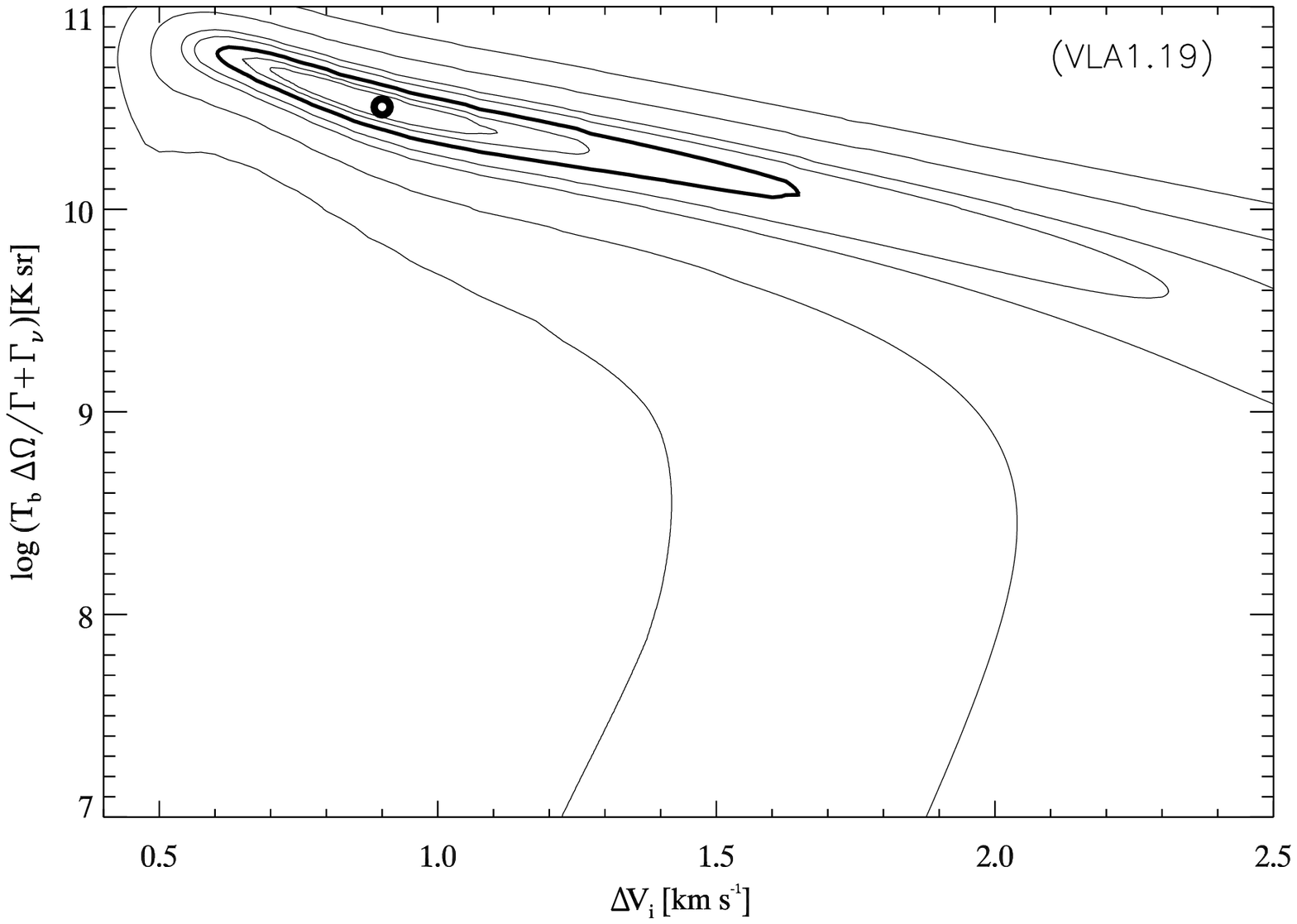}
\includegraphics[width = 8 cm]{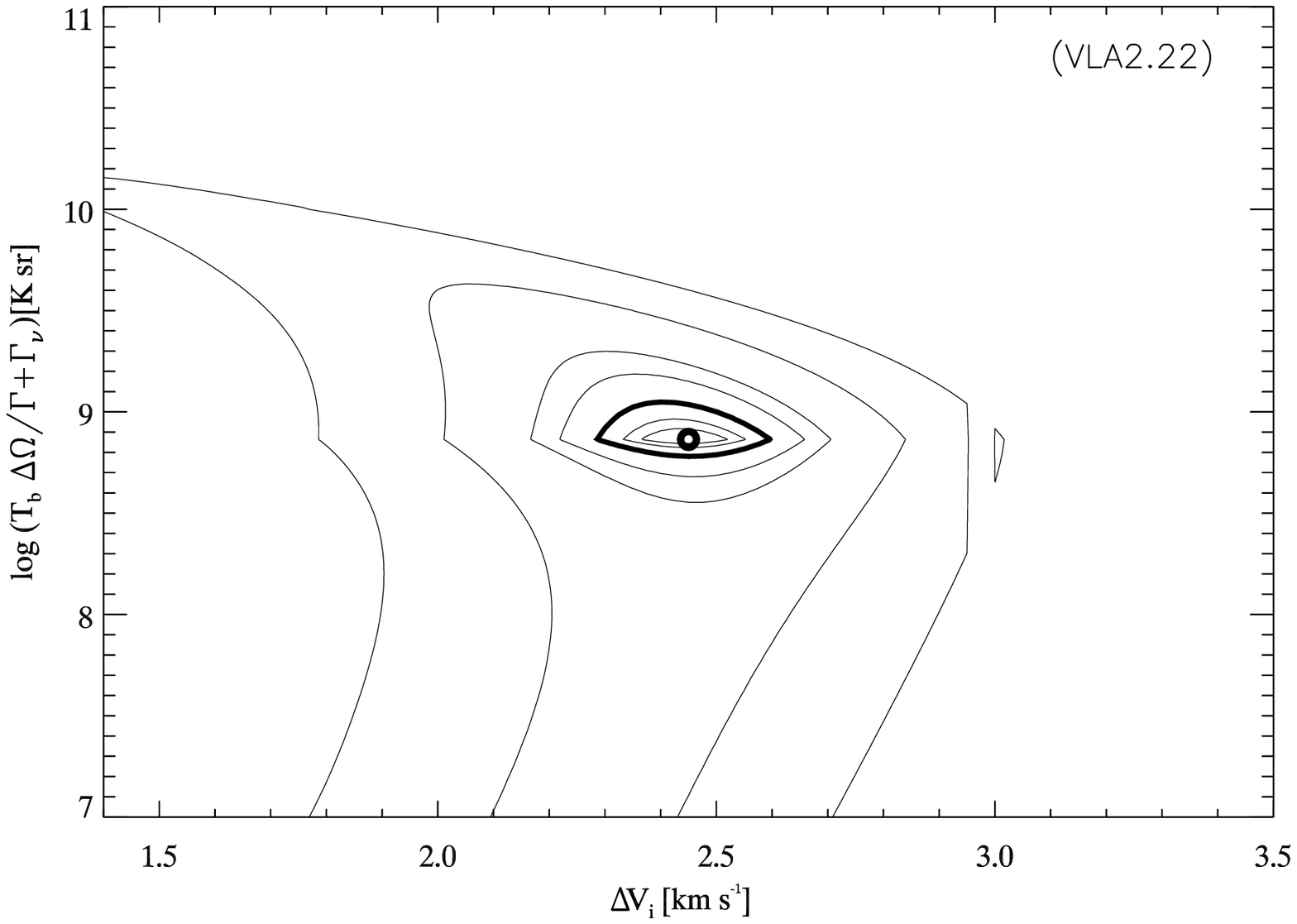}
\caption{Results of the full radiative transfer $\chi^{2}$-model fits, for the maser that shows the highest linear polarization fraction 
(VLA1.19) and for the maser that shows the highest $P_{\rm{V}}$ around VLA2 (VLA2.22). The fits yield the emerging maser brightness 
temperature $T_{\rm{b}}\Delta\Omega$ and the intrinsic maser linewidth $\Delta V_{\rm{i}}$. Contours indicate the significance intervals 
$\Delta \chi^{2}$=0.25, 0.5, 1, 2, 3, 7, with the thick solid contours indicating 1$\sigma$ and 3$\sigma$ areas.}
\label{idl}
\end{figure*}
\begin{figure*}[th!]
\centering
\includegraphics[width = 6 cm]{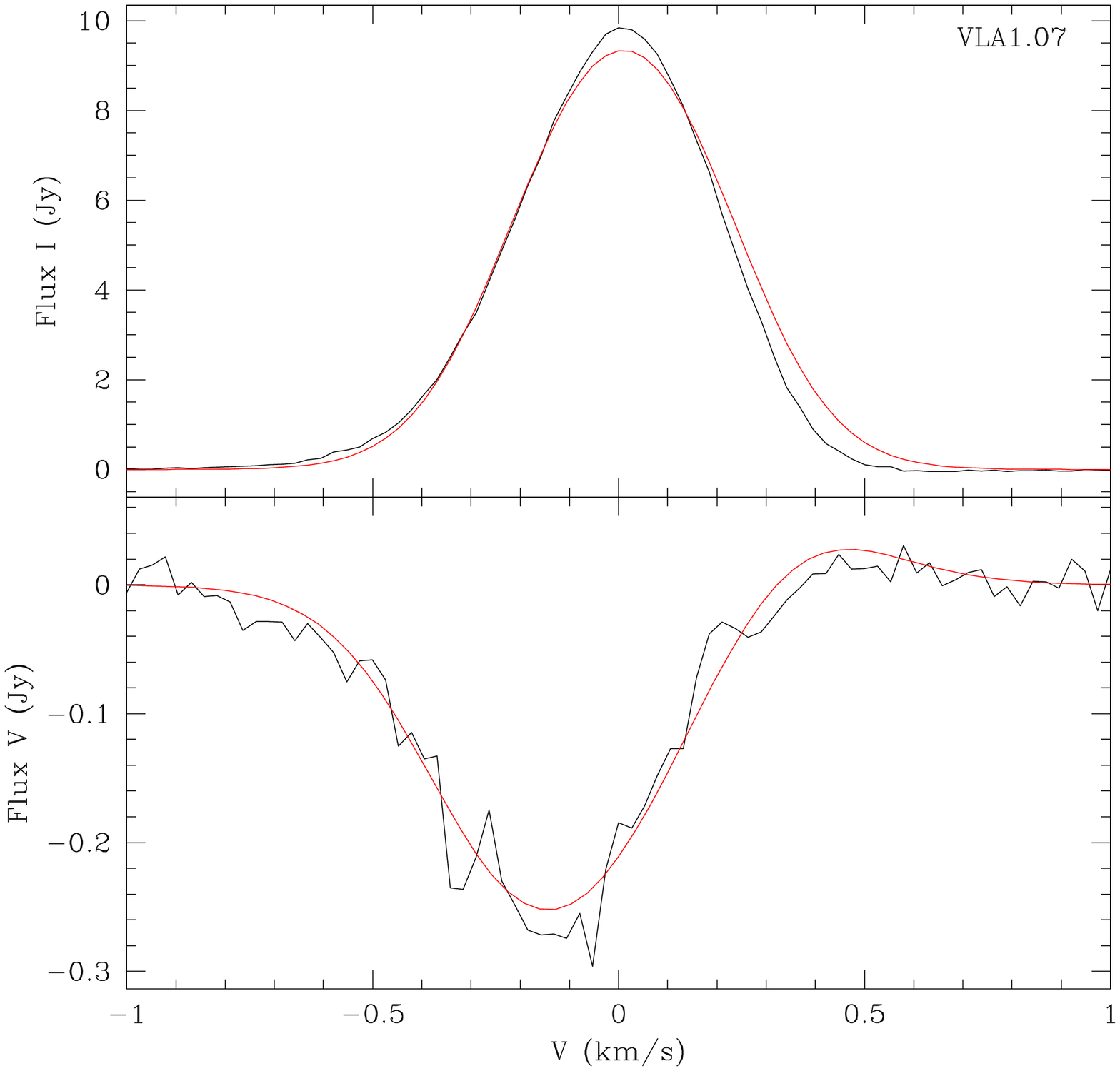}
\includegraphics[width = 6 cm]{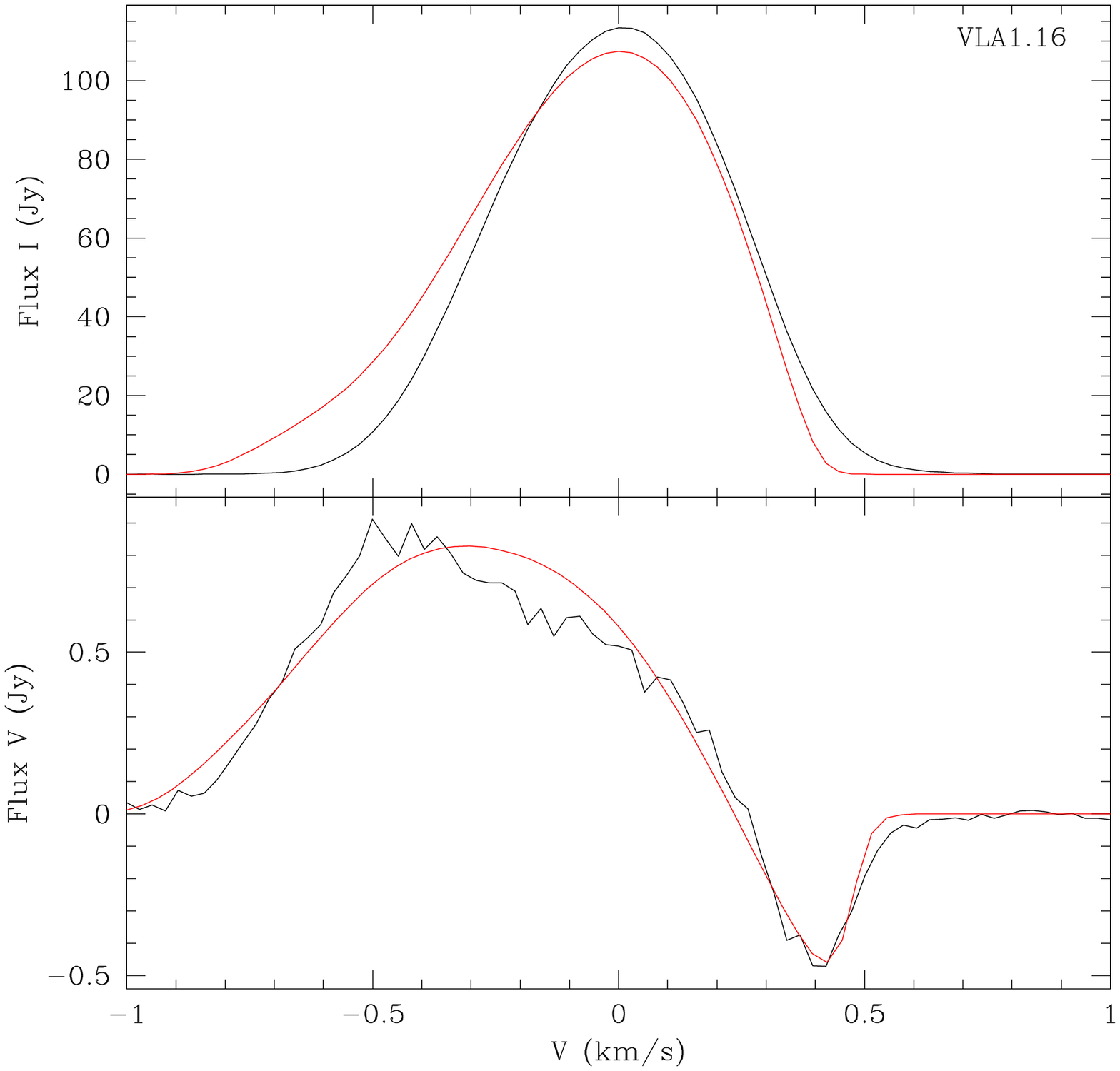}
\includegraphics[width = 6 cm]{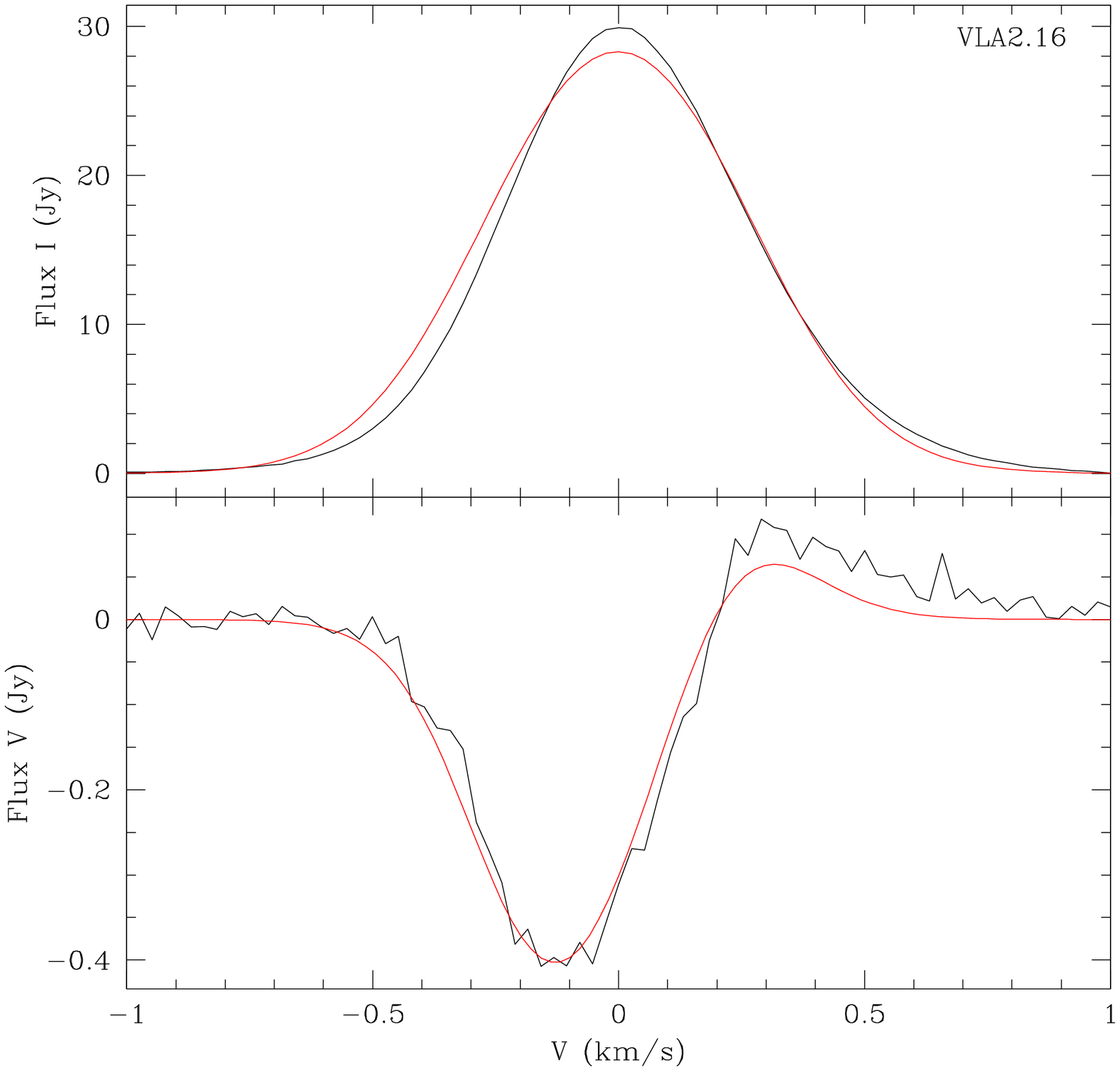}
\caption{Total intensity and circular polarization spectrum for the three maser features VLA1.07, VLA1.16, and VLA2.16. The thick red 
lines are the best-fit models of I and V emission obtained using the radiative transfer method (see Sect.~\ref{polfit}). All maser features
 were centered to zero velocity.}
\label{V}
\end{figure*}
\begin {table*}[t]
\caption []{All 22 GHz water maser features in W75N for which linear and/or circular polarization was detected.} 
\begin{center}
\scriptsize
\begin{tabular}{ l c c c c c c c c c c c c c }
\hline
\hline
\,\,\,\,\,(1)&(2)& (3)      & (4)      & (5)        & (6)            & (7)                 & (8)          & (9)          & (10)                    & (11)                        & (12)         & (13)     & (14)\\
\hline
Maser    & group$^{a}$& RA  & Dec      & Peak flux  & $V\rm{_{LSR}}$ & $\Delta v\rm{_{L}}$ & $P_{\rm{l}}$ &  $\chi$      & $\Delta V_{\rm{i}}^{b}$ & $T_{\rm{b}}\Delta\Omega^{b}$& $P_{\rm{V}}$ & $B_{||}$ & $\theta^{c}$\\
         & 	 & offset   & offset   & Density(I) &                &                     &              &	           &                    &                         &                  &           &            \\ 
	 & 	 & (mas)    & (mas)    & (\jyb)     &  (\kms)        &  (\kms)             &  (\%)        &   (\d)       & (\kms)             & (log K sr)              & (x10$^{-3}$)     &  (mG)     &  (\d)         \\
\hline
VLA\,1.01& C 	 & -95.3612 & -303.062 & $2.76\pm0.13$  & 10.7       & 0.38 & $2.3\pm0.4$  &$+43.8\pm8.3$ & $1.1^{+0.3}_{-0.3}$& $9.5^{+0.7}_{-0.8}$     &	$-$	     &	$-$	 & $79^{+10}_{-14}$ \\
VLA\,1.02& C 	 & -94.2666 & -300.461 & $3.39\pm0.09$  & 10.5       & 0.40 & $3.2\pm0.5$  &$+26.4\pm9.4$ & $1.1^{+0.3}_{-0.2}$& $9.6^{+0.6}_{-1.1}$     &	$-$	     &	$-$	 & $86^{+2}_{-13}$  \\
VLA\,1.03& C 	 & -91.6984 & -289.684 & $30.78\pm0.19$ & 11.3       & 0.39 & $5.1\pm0.8$  &$+29\pm1.6$   & $0.8^{+0.4}_{-0.3}$& $10.3^{+0.4}_{-0.3}$    &	$7.8$	     &$106\pm18$ & $74^{+10}_{-6}$  \\
VLA\,1.04& C	 & -72.0367 & -217.556 & $9.77\pm0.13$  & 10.7       & 0.46 & $3.9\pm0.4$  &$-43.0\pm1.4$ & $1.5^{+0.3}_{-0.4}$& $9.7^{+0.3}_{-0.3}$     &	$11.3$	     &$-212\pm51$& $83^{+5}_{-8}$   \\
VLA\,1.05& C	 & -71.5315 & -214.531 &$115.41\pm0.12$ & 10.6       & 0.44 & $1.7\pm0.2$  &$-56.1\pm6.5$ & $1.7^{+0.2}_{-0.3}$& $9.3^{+0.5}_{-0.1}$     &	$2.9$	     &$-54\pm9$	 & $74^{+16}_{-14}$ \\
VLA\,1.06& C 	 & -71.3210 & -213.371 & $76.79\pm0.12$ & 10.8       & 0.46 & $1.5\pm0.2$  &$-59.3\pm6.0$ & $1.8^{+0.2}_{-0.5}$& $9.4^{+0.6}_{-0.2}$     &	$5.1$	     &$-115\pm18$ & $66^{+11}_{-39}$ \\
VLA\,1.07& C	 & -70.4789 & -210.289 & $10.16\pm0.12$ & 11.1       & 0.45 & $1.4\pm0.6$  &$-53.9\pm11.3$& $1.9^{+0.2}_{-0.3}$& $9.2^{+0.6}_{-2.0}$     &	$28.4$	     &$599\pm100$& $72^{+8}_{-47}$  \\
VLA\,1.08& C	 & -69.5948 & -206.944 & $158.28\pm0.20$& 11.5       & 0.51 & $2.6\pm0.1$  &$-73.5\pm5.0$ & $1.8^{+0.2}_{-0.6}$& $9.7^{+0.5}_{-0.1}$     &	$2.8$	     &$65\pm10	$& $69^{+20}_{-5}$  \\
VLA\,1.09& C	 & -11.7886 & -121.338 & $6.14\pm0.06$  & 10.4       & 0.33 & $1.1\pm0.5$  &$-87.2\pm3.2$ & $1.0^{+0.2}_{-0.1}$& $9.0^{+0.2}_{-2.0}$     &	$-$	     &	$-$	 & $86^{+3}_{-52}$  \\
VLA\,1.10& C	 & -11.4097 & -237.705 & $2.07\pm0.20$  & 11.4       & 0.51 & $2.5\pm0.4$  &$-77.9\pm7.2$ & $2.0^{+0.2}_{-0.3}$& $9.5^{+0.5}_{-1.5}$     &	$-$	     &	$-$	 & $86^{+4}_{-4}$   \\
VLA\,1.11& C	 &  -9.7679 & -119.442 & $0.39\pm0.03$  & 10.3       & 0.51 & $2.7\pm0.3$  &$-56.1\pm14.2$& $1.0^{+0.2}_{-0.2}$& $9.5^{+0.6}_{-0.5}$     &	$-$	     &	$-$	 & $83^{+7}_{-9}$   \\
VLA\,1.12$^{d}$& B& -2.7366 & -15.713  & $1.13\pm0.03$  & 12.8       & 0.38 & $19.4\pm0.7$ &$-86.2\pm0.7$ & $0.7^{+0.3}_{-0.1}$& $10.5^{+0.1}_{-0.5}$    &	$-$	     &	$-$	 & $90^{+2}_{-2}$   \\
VLA\,1.13$^{d}$& B& -2.6945 & -14.427  & $0.97\pm0.03$  & 12.8       & 0.40 & $12.4\pm2.5$ &$-50.0\pm1.4$ & $0.8^{+0.2}_{-0.5}$& $10.1^{+0.3}_{-2.5}$    &	$-$	     &	$-$	 & $90^{+10}_{-10}$ \\
VLA\,1.14$^{d}$& B& -2.6945 & 1.568    & $80.02\pm0.07$ & 12.6       & 0.44 & $13.5\pm0.7$ &$-16.9\pm0.6$ & $0.6^{+0.2}_{-0.2}$& $10.6^{+0.1}_{-0.4}$    &	$-$	     &	$-$	 & $90^{+5}_{-5}$   \\
VLA\,1.15$^{d}$& B& -0.5894 & -36.270  & $63.25\pm0.17$ & 11.6       & 0.49 & $13.1\pm1.9$ &$-90.0\pm3.6$ & $0.7^{+0.8}_{-0.2}$& $10.7^{+0.1}_{-0.6}$    &	$-$	     &	$-$	 & $90^{+7}_{-7}$   \\
VLA\,1.16$^{d}$& B& -0.4210 & -35.545  & $111.47\pm0.20$& 11.4       & 0.55 & $12.9\pm1.2$ &$-66.9\pm0.6$ & $0.7^{+0.7}_{-0.2}$& $10.7^{+0.4}_{-0.3}$    &	$11.3$	     &$-177\pm27$& $90^{+6}_{-6}$   \\
VLA\,1.17$^{d}$& B&  0      & 0        & $2.96\pm0.06$  & 12.7       & 0.54 & $22.9\pm5.1$ &$-73.7\pm3.3$ & $1.1^{e}$          & $10.4^{+0.6}_{-2.3}$    &	$42.2$	     &$809\pm182$& $90^{+11}_{-11}$ \\
VLA\,1.18$^{d}$& B& 8.2520  & 31.166   & $6.17\pm0.06$  & 12.4       & 0.52 & $22.0\pm5.0$ &$-89.7\pm1.2$ & $1.0^{e}$          & $10.3^{+0.5}_{-2.3}$    &	$-$	     &	$-$	 & $90^{+12}_{-12}$ \\
VLA\,1.19$^{d}$& B& 8.2941  & 33.016   & $11.39\pm0.07$ & 12.6       & 0.53 & $25.7\pm3.5$ &$+58.2\pm7.7$ & $0.9^{e}$          & $10.5^{+0.4}_{-1.1}$    &	$5.9$	     &$-92\pm30$ & $90^{+6}_{-6}$   \\
VLA\,1.20& C	 & 36.2920  & -230.717 & $1.86\pm0.02$  & 8.1        & 0.33 & $2.7\pm0.3$  &$+47.1\pm2.9$ & $0.9^{+0.2}_{-0.1}$& $9.5^{+0.3}_{-0.3}$     &	$-$	     &	$-$	 & $84^{+3}_{-11}$  \\
VLA\,1.21$^{d}$& B& 57.6378 & 17.914   & $3.56\pm0.15$  & 11.2       & 0.44 & $-$          &$-$           & $-$                & $-$                     &	$25.8^{f}$   &$-398\pm102^{e}$& $-$        \\
VLA\,1.22$^{d}$& B& 84.3726 & 42.854   & $6.71\pm0.05$  & 9.8        & 0.43 & $0.9\pm0.2$  &$-20.6\pm6.2$ & $1.6^{+0.1}_{-0.3}$& $8.9^{+0.6}_{-1.3}$     &	$-$	     &	$-$	 & $78^{+12}_{-33}$ \\
VLA\,1.23& A	 & 295.3883 & 281.223  & $39.41\pm0.05$ & 21.1       & 0.83 & $0.6\pm0.1$  &$-30.6\pm8.5$ & $-$                & $-$                     &	$-$	     &	$-$	 & $-$              \\
VLA\,1.24& A	 & 332.4381 & 316.208  & $2.00\pm0.04$  & 21.7       & 0.68 & $1.9\pm0.1$  &$-31.8\pm8.3$ & $-$                & $-$                     &	$-$	     &	$-$	 & $-$              \\
VLA\,1.25& A	 & 507.5828 & 121.956  & $1.33\pm0.03$  & 23.8       & 0.51 & $3.6\pm0.3$  &$-9.9\pm7.6$  & $1.5^{+0.3}_{-0.4}$& $9.7^{+0.3}_{-1.0}$     &	$-$	     &	$-$	 & $90^{+11}_{-11}$ \\
VLA\,2.01& -	 & 571.4096 & -488.743 & $1.69\pm0.01$  & 5.2        & 0.57 & $-$          &$-$           & $-$                & $-$                     &	$30.2^{f}$   &$567\pm243^{e}$& $-$        \\
VLA\,2.02& -	 & 571.9570 & -487.320 & $2.82\pm0.01$  & 5.0        & 0.58 & $-$          &$-$           & $-$                & $-$                     &	$21.5^{f}$   &$413\pm146^{e}$& $-$        \\
VLA\,2.03& -	 & 587.4926 & -621.986 & $12.0\pm0.05$  & 6.8        & 0.68 & $0.5\pm0.1$  &$-89.5\pm52.7$& $3.0^{+0.2}_{-0.2}$& $8.7^{+0.5}_{-0.1}$     &	$-$	     &	$-$	 & $80^{+10}_{-36}$ \\
VLA\,2.04& -	 & 602.6914 & -611.675 & $6.86\pm0.06$  & 12.7       & 0.61 & $0.8\pm0.3$  &$+82.8\pm3.8$ & $-$                & $-$                     &	$-$	     &	$-$	 & $-$              \\
VLA\,2.05& -	 & 604.3755 & -620.571 & $4.62\pm0.12$  & 10.9       & 0.50 & $1.5\pm0.4$  &$-80.6\pm32.0$& $1.9^{+0.2}_{-0.3}$& $9.2^{+0.1}_{-2.3}$     &	$-$	     &	$-$	 & $90^{+22}_{-22}$ \\
VLA\,1.26& A	 & 605.6807 & 212.257  & $20.73\pm0.04$ & 18.7       & 0.53 & $2.3\pm0.1$  &$-17.9\pm2.0$ & $2.0^{+0.2}_{-0.4}$& $9.4^{+0.3}_{-0.1}$     &	$-$	     &	$-$	 & $85^{+3}_{-12}$  \\
VLA\,2.06& -	 & 625.2582 & -454.392 & $25.44\pm0.16$ & 11.2       & 0.69 & $0.2\pm0.1$  &$-47.0\pm16.6$& $-$                & $-$                     &	$-$	     &	$-$	 & $-$              \\
VLA\,2.07& -	 & 676.4543 & -432.327 & $1.95\pm0.08$  & 6.0        & 0.60 & $2.0\pm0.3$  &$+47.7\pm8.7$ & $2.3^{+0.2}_{-0.2}$& $9.4^{+0.2}_{-1.9}$     &	$-$	     &	$-$	 & $90^{+15}_{-15}$ \\
VLA\,2.08& -	 & 677.0438 & -432.545 & $4.65\pm0.01$  & 6.2        & 0.59 & $1.1\pm0.1$  &$+34.5\pm3.1$ & $2.5^{+0.2}_{-0.2}$& $9.0^{+0.3}_{-0.1}$     &	$-$	     &	$-$	 & $83^{+8}_{-16}$  \\
VLA\,2.09& -	 & 690.0533 & -439.411 & $30.29\pm0.06$ & 9.6        & 0.75 & $0.4\pm0.1$  &$-58.7\pm9.1$ & $3.3^{+0.1}_{-0.2}$& $8.6^{+0.7}_{-0.1}$     &	$3.9$	     &$-158\pm36$& $76^{+11}_{-40}$ \\
VLA\,2.10& -	 & 690.7269 & -440.163 & $48.76\pm0.06$ & 9.7        & 0.58 & $0.4\pm0.1$  &$-69.3\pm9.0$ & $2.6^{+0.1}_{-0.2}$& $8.6^{+0.8}_{-0.1}$     &	$6.2$	     &$-186\pm31$& $72^{+8}_{-44}$  \\
VLA\,2.11& -	 & 697.1685 & -459.564 & $13.99\pm0.04$ & -1.1       & 0.86 & $1.0\pm0.3$  &$-39.2\pm19.2$& $-$                & $-$                     &	$-$	     &	$-$	 & $-$              \\
VLA\,2.12& -	 & 697.4212 & -460.125 & $3.25\pm0.01$  & -2.9       & 2.34 & $1.5\pm0.6$  &$-34.2\pm12.1$& $-$                & $-$                     &	$-$	     &	$-$	 & $-$              \\
VLA\,2.13& -	 & 697.8843 & -458.710 & $7.16\pm0.01$  & 0.5        & 0.69 & $1.0\pm0.2$  &$-38.3\pm7.5$ & $3.0^{+0.1}_{-0.3}$& $9.0^{+0.6}_{-1.1}$     &	$-$	     &	$-$	 & $82^{+9}_{-13}$  \\
VLA\,2.14& -	 & 698.6000 & -445.961 & $62.39\pm0.07$ & 7.2        & 0.63 & $1.7\pm0.2$  &$-84.8\pm1.2$ & $2.7^{+0.3}_{-0.5}$& $9.3^{+0.7}_{-0.2}$     &	$1.1$	     &$-38\pm13$ & $74^{+15}_{-35}$ \\
VLA\,2.15& -	 & 698.9368 & -461.704 & $49.20\pm0.05$ & -0.7       & 0.87 & $2.0\pm0.2$  &$+78.2\pm36.9$&$3.4^{e}$          & $9.6^{+0.4}_{-0.3}$     &	$-$	     &	$-$	 & $68^{+3}_{-41}$  \\
VLA\,2.16& -	 & 699.3157 & -446.842 & $30.98\pm0.05$ & 6.8        & 0.58 & $1.3\pm0.2$  &$-75.4\pm1.6$ & $2.5^{+0.2}_{-0.3}$& $9.1^{+0.7}_{-0.3}$     &	$15.6$	     &$470\pm73$ & $77^{+13}_{-33}$ \\
VLA\,2.17& -	 & 701.5051 & -448.017 & $12.73\pm0.05$ & 8.3        & 0.53 & $2.5\pm0.5$  &$-71.7\pm1.7$ & $2.1^{+0.3}_{-0.3}$& $9.5^{+0.8}_{-0.9}$     &	$16.1$	     &$425\pm76$ & $77^{+13}_{-34}$ \\
VLA\,2.18& -	 & 702.3892 & -449.426 & $4.32\pm0.03$  & 9.3        & 0.82 & $1.2\pm0.1$  &$-76.9\pm4.4$ & $-$                & $-$                     &	$-$	     &	$-$	 & $-$              \\
VLA\,2.19& -	 & 703.4839 & -451.126 & $39.57\pm0.17$ & 11.6       & 0.78 & $2.6\pm0.3$  &$-72.6\pm3.7$ & $-$                & $-$                     &	$-$	     &	$-$	 & $-$              \\
VLA\,2.20& -	 & 703.8628 & -451.496 & $72.56\pm0.16$ & 11.2       & 1.09 & $2.0\pm0.3$  &$-65.1\pm3.9$ & $3.1^{+0.1}_{-0.9}$& $10.2^{+0.3}_{-0.2}$    &	$-$	     &	$-$	 & $61^{+42}_{-42}$ \\
VLA\,2.21& -	 & 705.5048 & -458.549 & $2.48\pm0.01$  & 0.2        & 0.53 & $1.4\pm0.3$  &$-8.7\pm5.2$  & $2.1^{+0.1}_{-0.3}$& $9.2^{+0.2}_{-2.0}$     &	$-$	     &	$-$	 & $90^{+16}_{-16}$ \\
VLA\,2.22& -	 & 709.9675 & -458.458 & $4.02\pm0.01$  & 5.0        & 0.56 & $0.8\pm0.1$  &$-8.0\pm4.4$  & $2.5^{+0.1}_{-0.2}$& $8.9^{+0.5}_{-0.4}$     &	$20.8$	     &$957\pm239$& $84^{+6}_{-11}$  \\
VLA\,2.23& -	 & 710.3886 & -459.141 & $1.34\pm0.01$  & 4.9        & 0.80 & $2.1\pm0.1$  &$-53.5\pm5.5$ & $3.0^{+0.2}_{-0.1}$& $9.4^{+0.1}_{-0.1}$     &	$-$	     &	$-$	 & $87^{+1}_{-9}$   \\
VLA\,2.24& -	 & 711.0201 & -457.901 & $198.45\pm0.20$& 8.7        & 0.51 & $0.7\pm0.1$  &$-72.0\pm6.2$ & $2.3^{+0.2}_{-0.3}$& $9.0^{+0.5}_{-0.3}$     &	$6.1$	     &$160\pm24	$& $66^{+11}_{-42}$ \\
VLA\,2.25& -	 & 713.0410 & -466.595 & $16.74\pm0.02$ & 13.7       & 0.60 & $0.8\pm0.1$  &$-44.5\pm2.1$ & $2.6^{+0.2}_{-0.3}$& $8.9^{+0.5}_{-0.1}$     &	$-$	     &	$-$	 & $80^{+10}_{-38}$ \\
VLA\,2.26& -	 & 716.0303 & -475.323 & $53.72\pm0.06$ & -7.7       & 0.68 & $1.2\pm0.5$  &$-89.6\pm61.8$& $3.0^{+0.2}_{-0.3}$& $9.1^{+0.5}_{-1.8}$     &	$2.1$	     &$79\pm18	$& $77^{+13}_{-38}$ \\
VLA\,2.27& -	 & 777.7941 & -447.278 & $0.67\pm0.06$  & -7.7       & 0.66 & $6.1\pm0.1$  &$-87.7\pm6.5$ & $0.7^{+1.8}_{-0.1}$& $10.7^{+0.1}_{-1.7}$    &	$-$	     &	$-$	 & $71^{+13}_{-3}$  \\
\hline
\end{tabular}
\end{center}
\scriptsize{$^{a}$ The water masers associated with VLA\,1 are divided in three groups according to their positions 
(see Sect.~\ref{masd}).\\ 
$^{b}$ The best-fitting results obtained by using a model based on the radiative transfer theory of water masers (Vlemmings et al. 
\cite{vle06}, \cite{vle10}) for $\Gamma+\Gamma_{\nu}=1$. The errors were determined by analysing the full probability distribution 
function. For $T\sim400$~K ($\Gamma_{\nu}=2$) and $T\sim2500$~K ($\Gamma_{\nu}=13$) $T_{\rm{b}}\Delta\Omega$ has to be adjusted by 
adding +0.48 and +1.15, respectively (Nedoluha \& Watson \cite{ned92}, Anderson \& Watson \cite{and93}).\\
$^{c}$The angle between the magnetic field and the maser propagation direction is determined by using the observed $P_{\rm{l}}$ and 
the fitted emerging brightness temperature.The errors were determined by analysing the full probability distribution function.\\
$^{d}$Because of the degree of the saturation of these water masers $T_{\rm{b}}\Delta\Omega$ is underestimated, $\Delta V_{\rm{i}}$ 
and $\theta$ are overestimated.\\
$^{e}$The constraint fit did not allow us to evaluate the error bars properly.\\
$^{f}$In the fitting model we include the values for $T_{\rm{b}}\Delta\Omega$ and $\Delta V_{\rm{i}}$ of the closest features.}
\label{polt}
\end{table*}
\subsection{Polarization fitting}
\label{polfit}
\indent From the maser theory we know that the fractional linear polarization $P_{\rm{l}}$ of the \hdueo \, maser emission depends on the
 degree of its saturation and the angle ($\theta$) between the maser propagation direction and the magnetic field (e.g., Goldreich 
et al. \cite{gol73}, Nedoluha \& Watson \cite{ned92}). 
Even the relation  between the measured polarization angle $\chi$ and the magnetic field angle on the sky ($\phi_{B}$) 
depends on $\theta$, with the linear polarization vector perpendicular to the field for $\theta>\theta_{\rm{crit}}\sim 55$\d \, (Goldreich 
et al. \cite{gol73}), where $\theta_{\rm{crit}}$ is the so-called Van Vleck angle. It is therefore important to evaluate $\theta$ for every 
\hdueo ~maser that shows linear polarization emission. In order to obtain the best values for $\theta$, we needed to determine the 
emerging brightness temperature ($T_{\rm{b}}\Delta\Omega$) and the intrinsic thermal linewidth ($\Delta V_{\rm{i}}$), 
which is the full width half-maximum (FWHM) of the Maxwellian distribution of particle velocities, of the masers with high 
precision. This could be done by fitting the water maser emissions with the full radiative transfer method described by Vlemmings et al. 
(\cite{vle06}, \cite{vle10}), who successfully used it for water and methanol masers in Cepheus A. These radiative transfer methods are 
based on the models for water masers of Nedoluha \& Watson (\cite{ned92}), for which the shapes of the total intensity,
 linear polarization, and circular polarization spectra depend on $T_{\rm{b}}\Delta\Omega$ and $\Delta V_{\rm{i}}$. They also 
found that the emerging brightness temperature scaled linearly with ($\Gamma+\Gamma_{\nu}$), which are the maser decay rate $\Gamma$ 
and cross-relaxation rate $\Gamma_{\nu}$. Since for the 
22~GHz \hdueo \, masers $\Gamma\lesssim 1 \rm{s^{-1}}$ and $\Gamma_{\nu}$ depends on the gas temperature, we include in our fit a value 
$(\Gamma+\Gamma_{\nu})=1 \rm{s^{-1}}$, so that we could adjust the fitted $T_{\rm{b}}\Delta\Omega$ values by simply scaling according to
 ($\Gamma+\Gamma_{\nu}$). Note that $\Delta V_{\rm{i}}$ and $\theta$ do not need to be adjusted.\\
\indent We model the observed linear polarized and total intensity maser spectra by 
gridding the intrinsic thermal linewidth $\Delta V_{i}$ between 0.4 and 3.5~\kms, in steps of 0.025~\kms, using a least square 
fitting routine ($\chi^{2}$-model) with an upper limit of the brightness temperature 
$T_{\rm{b}}\Delta\Omega=10^{11}$~K~sr. In Fig.~\ref{idl} two examples are shown of the results obtained by fitting the water 
maser emissions with the full radiative transfer method code. From these first fits, we can then obtain the angles $\theta$ by 
considering the relation between $P_{\rm{l}}$ and $\theta$ (see Vlemmings et al. \cite{vle10} for more details). The best values for 
$T_{\rm{b}}\Delta\Omega$ and $\Delta V_{\rm{i}}$ are then included in the full radiative transfer code to produce the 
\textit{I} and \textit{V} models that were used for fitting the total intensity and circular polarized spectra of the \hdueo 
~masers extracted from the high-resolution \textit{I} and \textit{V} cubes. The magnetic field strength along the line of sight is 
evaluated by using the equation
\begin{equation}
B_{||}=B~cos \theta= \frac{P_{V}~\Delta v_{L}}{2 \cdot A_{F-F'}},
\label{magn}
\end{equation}
where $\Delta v_{L}$ is the FWHM of the total intensity spectrum, the $A_{\rm{F-F'}}$ coefficient, which
 depends on $T_{\rm{b}}\Delta\Omega$. The circular polarization fraction ($P_{\rm{V}}$) 
are obtained from the $I$ and $V$ models.\\
\begin{figure*}[th!]
\centering
\includegraphics[width = 13 cm, angle = -90]{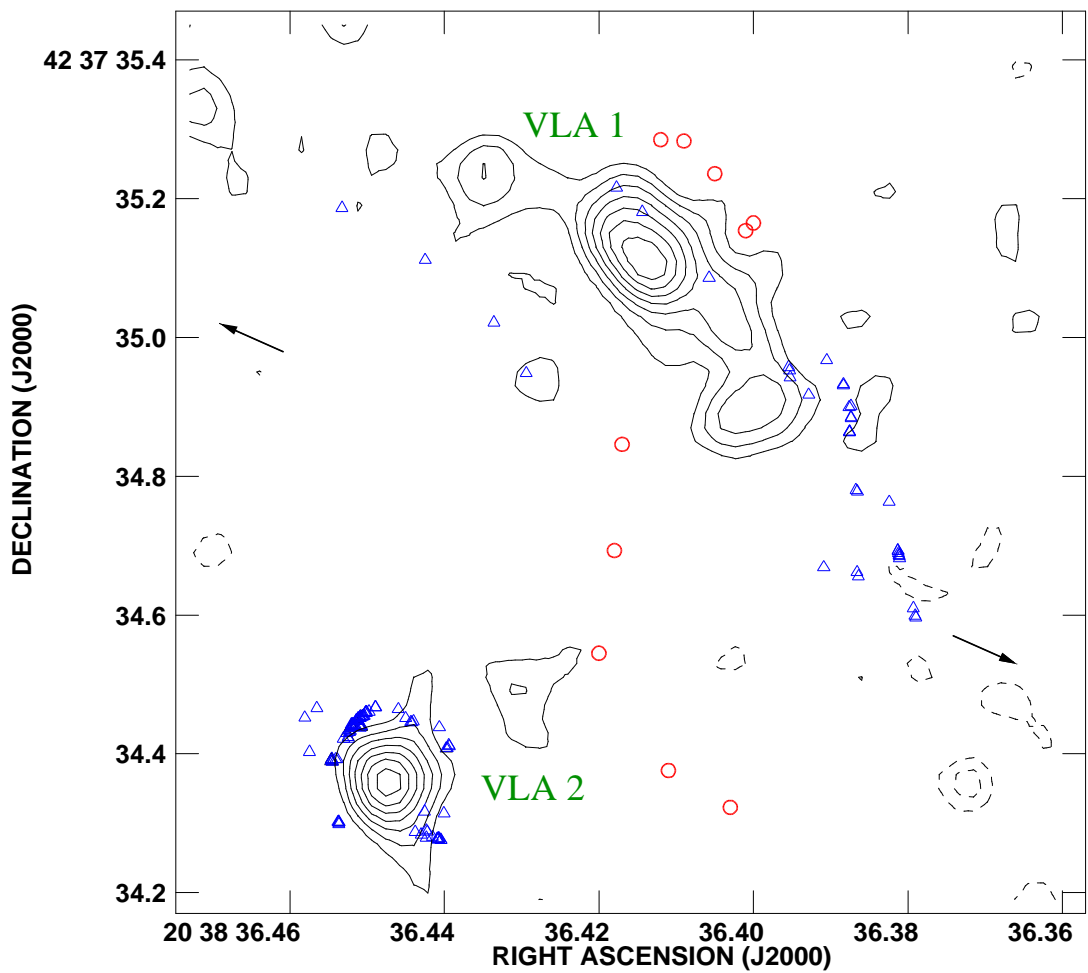}
\caption{Positions of water and methanol masers superimposed on the 1.3 cm continuum contour map of the VLA1 thermal jet and VLA\,2 
(Torrelles et al. \cite{tor97}). Contours are $-3$, $-2$, 2, 3, 4, 5, 6, 7, 8 $\times$ 0.16~\mjyb. Blue triangles indicate the positions 
of the water maser features detected with the VLBA (this paper). Red circles indicate the position of the methanol masers features 
detected by S09. The two arrows indicate the direction of the large-scale molecular bipolar outflow (PA=66\d).}
\label{pos}
\end{figure*}
\indent Another important aspect that must be considered in the analysis of the water masers is the degree of their saturation, which 
influences $P_{\rm{l}}$. We are able to estimate it by considering the ratio between maser rate of stimulated emission ($R$) and the sum 
of maser decay rate and the cross-relaxation rate ($\Gamma+\Gamma_{\nu}$).\\
\indent The stimulated emission rate is given by
\begin{equation}
R\simeq\frac{A k_{\rm{B}} T_{\rm{b}}\Delta\Omega}{4\pi h \nu},
\label{sateq}
\end{equation}
where $A=2\times10^{-9}\,\rm{s^{-1}}$ is the 22-GHz \hdueo \, maser spontaneous rate (Goldreich \& Keeley \cite{gol72}), $k_{\rm{B}}$ and 
$h$ are the Boltzmann and Planck constants, respectively, $\nu$ the maser frequency and $T_{\rm{b}}\Delta\Omega$ the emerging brightness 
temperature adjusted at the right $\Gamma+\Gamma_{\nu}$ value. The masers are unsaturated when $R/(\Gamma+\Gamma_{\nu})<1$, in the onset 
of saturation when $R/(\Gamma+\Gamma_{\nu})\geq1$,  and fully saturated when $R/(\Gamma+\Gamma_{\nu})\approx100$ (Vlemmings et al. 
\cite{vle06}). Moreover, when the saturation sets in, the maser lines start to broaden again, and this means that the observed linewidth 
($\Delta v\rm{_{L}}$) becomes as large as $\Delta V_{\rm{i}}$ or larger  depending on the degree of the saturation.\\
\indent Although the full radiative transfer method is based on a model for general water masers, it should be noted 
that, when this model is applied to \textit{saturated masers}, it is impossible to properly disentangle the values of 
$T_{\rm{b}}\Delta\Omega$ and 
$\Delta V_{\rm{i}}$. In this case the method only provides a lower limit for $T_{\rm{b}}\Delta\Omega$ and an upper limit for 
$\Delta V_{\rm{i}}$. For high maser brightness temperatures, i.e. greater than 
$10^{9}\rm{cm^{-3}}$~K~sr (Nedoluha \& Watson \cite{ned92}), the $\rm{cos}\theta$ dependence of Eq.~\ref{magn} breaks down thereby 
introducing a more complex dependence on $\theta$. As a result, the angle $\theta$ obtained from the fit is overestimated and, in particular, 
it could also reach values of 90\d ~or greater. Consequently, the values of $T_{\rm{b}}\Delta\Omega$, $\Delta V_{\rm{i}}$ and 
$\theta$ obtained for saturated masers could not be taken into account in the analysis of the region.
\section{Results}
\subsection{Maser distribution}
\begin{figure*}[th!]
\centering
\includegraphics[width = 9 cm]{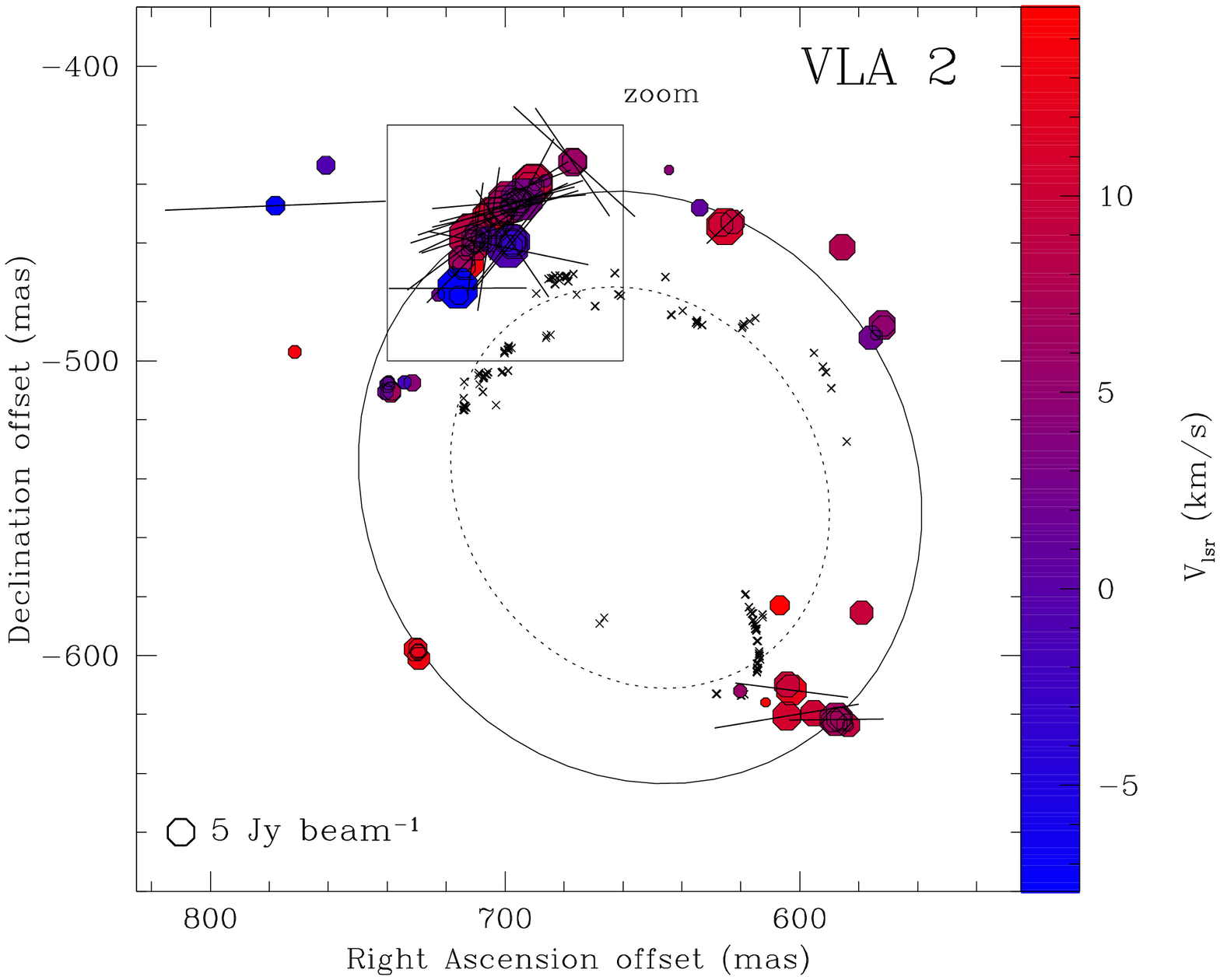}
\includegraphics[width = 9 cm]{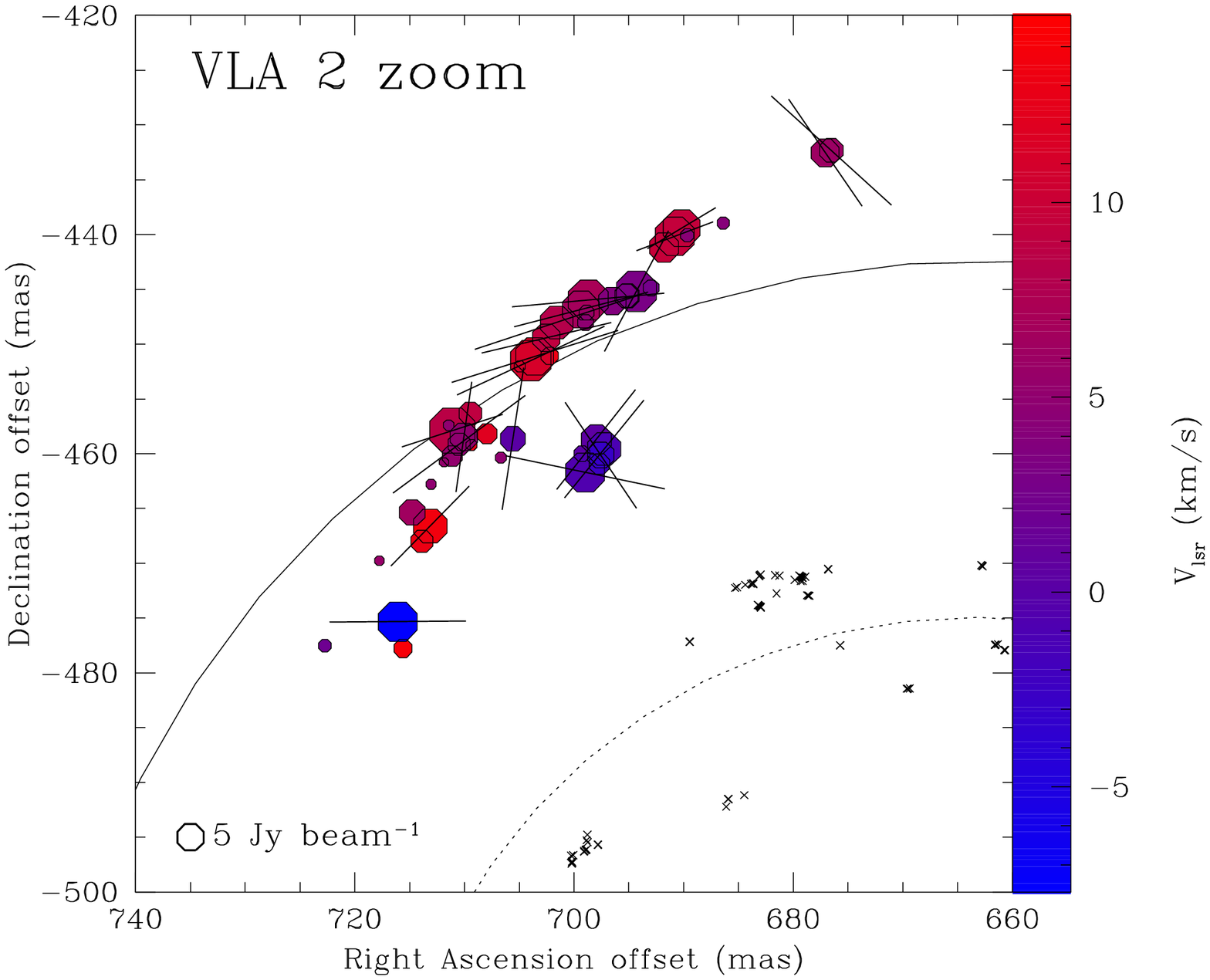}
\caption{Left panel: a close-up view of the \hdueo \, maser features around the radio source VLA\,2. Right panel: a zoom-in view of the 
boxed region of the left panel. The octagonal symbols are the identified maser features in present work scaled logarithmically according 
to their peak flux density. The maser LSR radial velocity is indicated by color. A 5~\jyb \, symbol is plotted for illustration in both 
panel. The linear polarization vectors, scaled logarithmically according to polarization fraction $P_{\rm{1}}$ (in Table~\ref{polt}), are 
overplotted. Two ellipses are also drawn in both panels. They are the results of the best fit of the water masers (crosses) detected by 
T03 (dotted ellipse, ellipse 1; epoch 1999) and of those detected in present work (solid ellipse, ellipse 2; epoch 2005). Their parameters
 are listed in Table~\ref{vla2}. The synthesized beam is 2.0 mas $\times$ 0.7 mas. }
\label{vla2}
\end{figure*}

In Fig.~\ref{pos} we show the distribution of the 124 22-GHz water maser features detected with the VLBA. No water maser emission with a 
peak flux density less than 80~\mjyb \, is detected. All \hdueo \, masers are detected around the radio sources VLA\,1 (29\%, 36/124) 
and VLA\,2 (71\%, 88/124).\\
\indent The water masers associated with VLA\,2, which have local standard of rest velocities ($V_{\rm{LSR}}$) between -7.7 and 13.7~\kms,
 are distributed elliptically around its 1.3\,cm continuum peak. In Fig.~\ref{vla2} we report two ellipses obtained by fitting the 
positions of the water masers detected by T03 (called here ellipse 1) and in the present paper (ellipse 2), their parameters are reported 
in Table~\ref{fitel}. The major axis of ellipse 2 is 64~mas larger than that of 
ellipse 1, the minor axes are also in the same ratio. This indicates an expansion velocity of $\sim$4.8~mas/yr in all directions, 
corresponding to $\sim$ 46~\kms \, (at 2~kpc). This value is consistent with the proper motion of the water masers as reported by T03. Since 
the expansion velocity is symmetric in all directions, we aligned the centers of the two ellipses with the position of the 1.3~cm 
continuum peak obtained from a Gaussian fit. Consequently we were able to overlay all water masers of the region on the continuum VLA map 
with an accuracy of approximately 10\,mas. In this way we could also compare the positions of the masers around VLA\,1 at the two 
different epochs. If we look at the northern masers around VLA\,2, we see that the masers detected by T03 are better aligned with the 
ellipse 1 than the masers detected by us with ellipse 2 (Fig.~\ref{vla2}, right panel), indicating that those masers are also moving 
northeastward. This movement might be due to the formation of a jet (see Fig.~\ref{pos}).\\
\indent The water masers associated with VLA\,1 can be divided into three groups, one composed of the northern masers (VLA\,1A), one of the 
masers closest to the central protostar (VLA\,1B, see%
\begin {table}[h]
\caption []{Number, epoch, semi-major axis, semi-minor axis, and inclination from the elliptical fit of the water maser associated with 
VLA\,2 in the two different epochs.} 
\begin{center}
\scriptsize
\begin{tabular}{c c c c c c}
\hline
\hline
Ellipse & epoch        & $a$    & $b$    & $\theta$ & ref.  \\
        &              & (mas)  & (mas)  & (\d)     &       \\
\hline
\\
1       & 2-Apr.-1999  & 71.0   & 61.1   & 123.5    & T03   \\
2       & 21-Nov.-2005 & 103.1  & 92.9   & 120.5    & this work\\
\\
\hline
\end{tabular}
\end{center}
\label{fitel}
\end{table}
\begin{figure*}[t!]
\centering
\includegraphics[width = 9 cm]{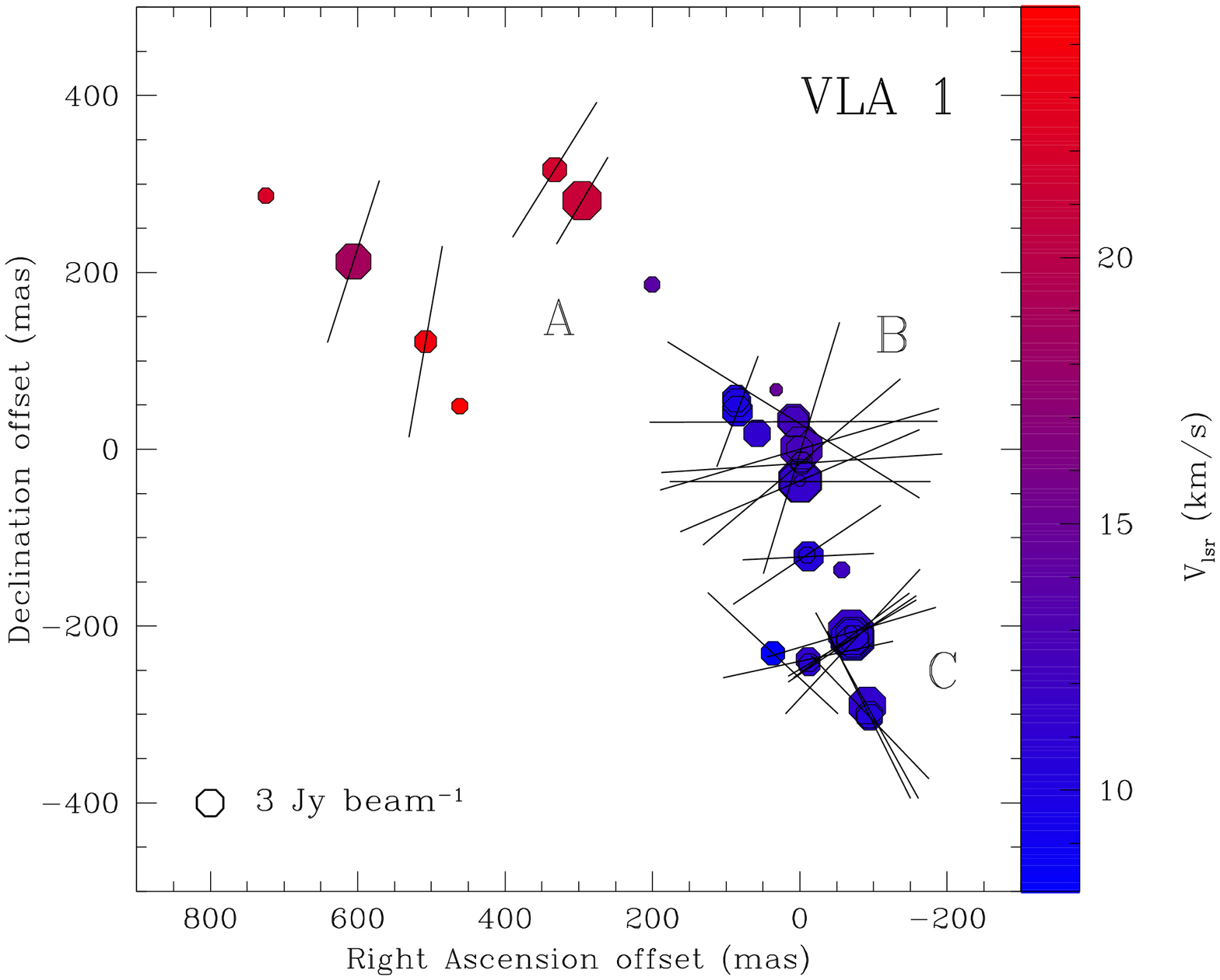}
\includegraphics[width = 9 cm]{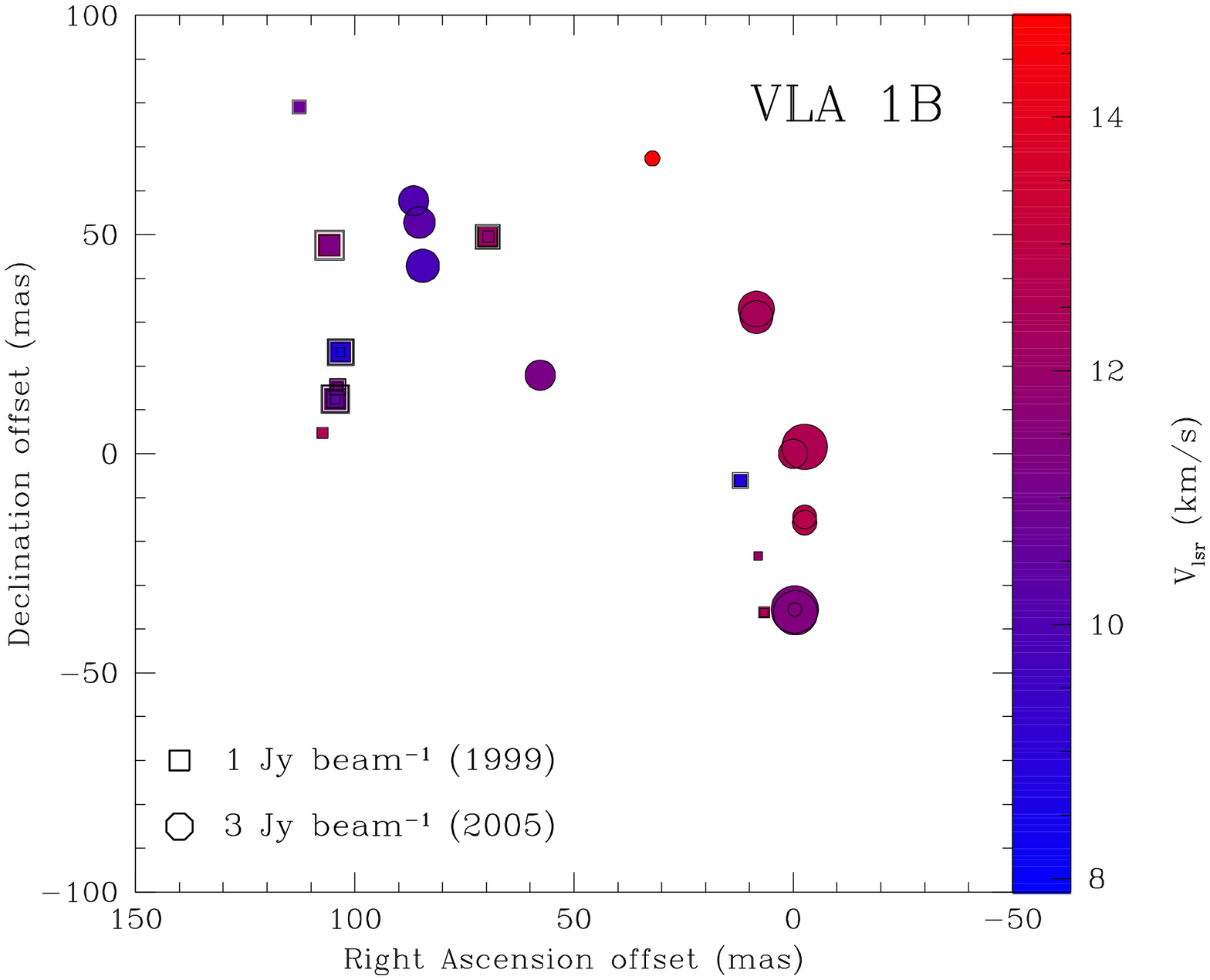}
\caption{Left panel: a close-up view of the \hdueo \, maser features (octagon) around the radio source VLA\,1. Right panel: a zoom-in 
view of the water masers group VLA\,1B, here are also reported the water masers (square) detected by T03. The octagonal and the square 
symbols are scaled logarithmically according to their peak flux density. The maser LSR radial velocity is indicated by color. The systemic
 LSR radial velocity of W75N massive star-forming region is 10~\kms \,(Shepherd et al. \cite{she03}). Symbols for 3~\jyb \, and for 
1~\jyb \, are shown in the lower left corner of the panels. The linear polarization vectors, scaled logarithmically according to 
polarization fraction $P_{\rm{1}}$ (in Table~\ref{polt}), are overplotted on the left panel. The synthesized beam is 2.0 mas 
$\times$ 0.7 mas.}
\label{vla1}
\end{figure*}
Fig.~\ref{vla1} right panel) and the third one of the southern masers (VLA\,1C). The east-northern masers of the VLA\,1A group  did not 
show emission during the previous observation of epoch 1999 (T03). All VLA\,1A masers ($V_{\rm{LSR}}\simeq 19-25$~\kms) are well-aligned 
with the direction of the thermal radio jet detected by Torrelles et al. (\cite{tor97}) and the methanol masers detected by S09. The 
VLA\,1B and VLA\,1C groups show a distribution quite similar to that shown on April 2$\rm{^{nd}}$ 1999 (T03). For these two groups T03 
reported a proper motion of about $\sim$2 mas/yr ($\sim19$~\kms). Although we have detected several water masers close to those detected 
by T03, it is very difficult to estimate the proper motion between the two epochs. In fact, as is shown on the right panel 
of Fig.~\ref{vla1}, it is difficult to identify a maser that arises in both epochs (1999 and 2005). The LSR radial velocities of VLA\,1 
are between 9 and 25~\kms, which are about 3-4 times higher than those of methanol masers (S09) and greater than the systemic velocity of 
the whole region (10~\kms, Shepherd et al. \cite{she03}).
\label{masd}
\subsection{Linear and circular polarization}
\indent Linear polarization was detected in 50 masers (Table~\ref{polt}). In particular, about 70\% of the \hdueo \, masers associated 
with VLA\,1 show linear polarization emission. The corresponding linear polarization vectors are shown in Figs.~\ref{vla2} and \ref{vla1} 
(left panel). The highest fractional linear polarization was detected in water masers related to VLA\,1 ($P_{\rm{l}}\sim26\%$). Eight 
masers associated with VLA\,1 have high linear polarizations, $P_{\rm{l}}$ between 10\% and 26\%, so that the median value of linear 
polarizations for the masers in VLA\,1 is $\sim$7\%. In contrast, the maximum $P_{\rm{l}}$ for the masers associated with VLA\,2 is 6\%, 
and the median value is much lower at 1.5\%. Circular polarization was detected in 20 masers, equally distributed between VLA\,1 and 
VLA\,2.\\
\indent The rms weighted linear polarization angles of the water masers around VLA\,1 and VLA\,2 are
 $\langle\chi_{1}\rangle\approx -67$\d$\pm40$\d \, and $\langle\chi_{2}\rangle\approx -72$\d$\pm32$\d, respectively. Since the water masers 
of group VLA\,1A show ordered linear polarization vectors, it is worthwhile evaluating its weighted polarization angles, which is 
$\langle\chi_{\rm{1A}}\rangle= -20$\d$\pm9$\d.\\
\indent We were able to fit 42 masers with the method described in \S~\ref{polfit}, and the results are given in columns 10 and 11 of 
Table~\ref{polt}, and two examples (i.e., VLA1.19, VLA2.22) are reported in Fig.~\ref{idl}. Although the emerging brightness temperatures 
for VLA\,1 and VLA\,2 are both in the range $10^{8}\,\rm{K \,sr}<$\,$T$\,$_{\rm{b}}\Delta\Omega<10^{10}\, \rm{K\,sr}$, their weighted 
intrinsic maser linewidths are very different. They are $\langle\Delta V_{\rm{i}}\rangle_{\rm{VLA1}}=1.0^{+0.8}_{-0.2}$~\kms \, and 
$\langle\Delta V_{\rm{i}}\rangle_{\rm{VLA2}}=2.5^{+0.6}_{-0.4}$~\kms\,, respectively.\\
\indent Finally, considering the emerging brightness temperature and the observed $P_{\rm{l}}$ we determined $\theta$ for those masers 
which we were able to fit for $\Delta V_{\rm{i}}$ and $T_{\rm{b}}\Delta\Omega$. The results are given in column 14 of Table~\ref{polt} with 
the errors determined by analyzing the full probability distribution functions. Eight of the \hdueo \, masers around the radio source VLA\,1 
(from VLA1.12 to VLA1.19, which are the closest features to the protostar) show values for $\theta$ equal to 90\d. This suggests that, 
from the theory of the Zeeman effect, we are observing perpendicular to the magnetic field. Consequently, we should not be able to detect 
circular polarization from these masers. However, since we detected circular polarization in three of them (VLA1.16, VLA1.17, and VLA1.19) 
the measurements of $\theta$ must be affected by the degree of their saturation, as discussed in \S~\ref{polfit}. 
By excluding these 8 masers, we evaluated $\langle\theta\rangle_{\rm{VLA1}}=83$\d$\pm7$\d \, and 
$\langle\theta\rangle_{\rm{VLA2}}=85$\d$\pm6$\d, which imply that the magnetic fields are close to the plane of the sky.\\
\indent In column 13 of Table~\ref{polt} we report the magnetic field strength along the line of sight obtained by fitting the 
high-resolution spectra of \textit{I} and \textit{V} (see \S~\ref{polfit}). The masers VLA1.21, VLA2.01, and VLA2.02 do not show any linear 
polarization emission, but they do show circular polarization emission. Consequently we could not get $T_{\rm{b}}\Delta\Omega$ and 
$\Delta V_{\rm{i}}$ as done for the other masers that showed linear polarization emission, and to measure the magnetic field 
strength, we used the $T_{\rm{b}}\Delta\Omega$ and $\Delta V_{\rm{i}}$ values of the closest masers. In Table~\ref{polt} the circular 
polarization fraction is also reported (column 12). 
\section{Discussion}
The water masers detected using the VLBA are related to two different radio continuum sources of W75N(B), VLA\,1 and VLA\,2, which are 
thought to be in two different evolutionary stages. In the next sections, we discuss their nature as two independent groups, their
 polarization and finally the type of shocks that give rise to them.
\subsection{\hdueo \, maser properties}
As suggested by the difference between their intrinsic thermal linewidths, which is 1.5~\kms, the masers around VLA\,1 and VLA\,2
 appear to arise under two different physical conditions. Using
\begin{equation}
\Delta V_{\rm{i}}\approx 0.5\times (T/100)^{\frac{1}{2}},
\label{tv}
\end{equation}
we estimate the gas temperatures in the \hdueo \, masing regions around VLA\,1 and VLA\,2, and find them to be $T_{\rm{VLA\,1}}\sim400$~K 
and $T_{\rm{VLA\,2}}\sim2500$~K, respectively. The intrinsic thermal linewidths of the VLA\,2 masers are quite similar to each
 other, while the three groups of masers around VLA\,1 show a difference in their $\Delta V_{\rm{i}}$. In particular, we find that the 
maser groups VLA\,1A and C have similar intrinsic thermal linewidths ($\langle\Delta V_{\rm{i}} \rangle=1.0$~\kms) while the
 group VLA\,1B has lower values ($\langle\Delta V_{\rm{i}} \rangle=0.7$~\kms). This can come from the degree of their saturation that we
 can evaluate by considering Eq.~\ref{sateq}.\\
\indent Before using Eq.~\ref{sateq} we have to adjust the emerging brightness temperature values as explained in \S~\ref{polfit}. At a 
temperature of $\sim$400\,K, for which $\Gamma+\Gamma_{\nu}=3\, \rm{s^{-1}}$ (Nedoluha \& Watson \cite{ned92}), the weighted emerging 
brightness temperature of VLA\,1B is $\langle T_{\rm{b}}\Delta\Omega\rangle>10^{11}$\,K\,sr, which is close to the limit of full 
saturation, and for VLA\,1A and C is $\langle T_{\rm{b}}\Delta\Omega\rangle=10^{10}$~K~sr.  From Eq.~\ref{sateq} the masers of 
VLA\,1A and VLA\,1C are thus completely unsaturated ($R/(\Gamma+\Gamma_{\nu})=0.5$), while VLA\,1B masers are saturated 
($R/(\Gamma+\Gamma_{\nu})>5$), even if the real degree of their saturation is unknown because of underestimating 
$T_{\rm{b}}\Delta\Omega$. The average observed linewidth ($\Delta v\rm{_{L}}$) of VLA\,1B group, which is about 0.5~\kms, gives us 
another indication of the saturation state of their \hdueo \, masers. As reported in \S~\ref{polfit}, the intrinsic thermal 
linewidths for the saturated masers are overestimated so $\Delta v\rm{_{L}}$ might be larger or at least equal to $\Delta V_{\rm{i}}$ 
implying that the maser lines are rebroadened as expected in the saturated maser lines. As the VLA\,1B masers are saturated, they 
have not been taken into account in our conclusions. For the other masers associated with VLA\,1, we observed linewidths
 well below their intrinsic thermal linewidth, and this confirms that these masers are unsaturated. The masers associated with 
VLA\,2 are unsaturated; in fact, for $T\sim2500$\,K ($\Gamma+\Gamma_{\nu}=14\, \rm{s^{-1}}$, Anderson \& Watson \cite{and93}), we 
get ratios of $R/(\Gamma+\Gamma_{\nu})<0.5$.\\
\indent From the measurements of the maser flux densities ($S(\nu)$) and feature angular sizes ($\Sigma$) 
we can estimate the brightness temperature by the equation reported in S09,
\begin{equation}
\frac{T_{\rm{b}}}{[\rm{K}]}=\frac{S(\nu)}{[\rm{Jy}]}\cdotp \left(\frac{\Sigma^2}{[\rm{mas^2}]}\right)^{-1}\cdotp\xi_{\rm{H_{2}O}}~,
\label{tb}
\end{equation}
where $\xi_{\rm{H_{2}O}}=1.24\times10^{9}\,\rm{mas^{2}\,Jy^{-1}\,K}$ is a constant factor which includes all constant values, such as the
 Boltzmann constant, the wavelength, and the proportionality factor obtained for a Gaussian shape by Burns et al. (\cite{bur79}). 
Comparing $T_{\rm{b}}$ with the emerging brightness temperatures obtained from the model, we can easily estimate $\Delta \Omega$. The 
Gaussian fit of the masers give a size of 0.4\,mas for the water masers related to VLA\,1, which indicates that they are unresolved, 
while a size of 0.5 for the masers associated with VLA\,2, which indicates that they are marginally resolved. The brightness temperatures
 of the brightest masers in both groups are $T_{\rm{b, VLA1.08}}>1.23\times10^{12}$~K and $T_{\rm{b, VLA2.24}}\approx10^{12}$~K, for 
which we find $\Delta \Omega_{\rm{VLA1.08}}<10^{-2}$~sr and $\Delta \Omega_{\rm{VLA2.24}}\approx10^{-2}$\,sr for $T=400$~K and 2500~K, 
respectively. Considering all the features in Table~\ref{polt}, the maser beaming of the two groups are  
$\Delta \Omega_{\rm{VLA\,1}}\lesssim10^{-2}$~sr ($\Gamma+\Gamma_{\nu}=2\,\rm{s}^{-1}$) and $\Delta \Omega_{\rm{VLA\,2}}\approx10^{-2}$~sr 
($\Gamma+\Gamma_{\nu}=14\,\rm{s}^{-1}$). In a tubular geometry $\Delta \Omega\approx(d/l)^{2}$, where $d$ and $l$ are the transverse size 
and length of the tube respectively. Assuming $d$ approximately the size of the maser features, the maser lengths are all in the range 
$10^{12}\,\rm{cm}\lesssim $~$l$~$\lesssim10^{14}\,\rm{cm}$.
\label{prop}
\subsection{Magnetic field in W75N(B)}
\label{polma}
\subsubsection{Magnetic field strength}
The magnetic field strength along the line of sight was determined from circular polarization measurements for 20 water masers. We 
detected significant ($\geq3\sigma$) magnetic fields toward 10 masers in VLA\,1 and 7 masers in VLA\,2. The detected fields along the 
line of sight ($B_{||}$) range from $-400$~mG to $+600$~mG for the \hdueo \, masers around VLA\,1 and from $-200$~mG to $1000$~mG for 
those associated with VLA\,2 (see Table~\ref{polt}). The wide range obtained for $B_{||}^{VLA1}$ and  $B_{||}^{VLA2}$ is not 
anomalous, and a similar range was reported for 
Cepheus\,A (Vlemmings et al. \cite{vle06}). The changing in the sign indicates the reversal of the magnetic field (negative towards the 
observer, positive away from the observer) as already observed in other sources (e.g., Vlemmings et al. \cite{vle06}). Since the $\theta$
 values are close to 90\d, i.e. close to the plane of the sky, a slight difference in these angles can produce an inversion of the sign 
of the magnetic fields. In the case of VLA\,1, the positive and negative magnetic field strengths measured for VLA\,1B and VLA\,1C groups 
indicate a twisted magnetic field, which is not the case for VLA\,2. Apart from the \hdueo \, masers VLA2.01 and VLA2.02, which are located 
on the right edge of the ellipse 2 and show positive values, the other masers in the upper left part of the ellipse 2 (from VLA2.09 to 
VLA2.26) present a clear separation between positive and negative magnetic field strengths (see Fig.~\ref{vla2} and Table~\ref{polt}).\\
\indent Based on these detections the absolute weighted magnetic field strengths 
along the line of sight, where the weights are $w_{\rm{i}}=1/e^{2}_{\rm{i}}$ and $e_{\rm{i}}$ is the error of the i-th measurement, are 
$\langle|B_{||}^{\rm{VLA1}}|\rangle=(81\pm62)$~mG and $\langle|B_{||}^{\rm{VLA2}}|\rangle=(145\pm110)$~mG. In the case 
of VLA\,1, it is more correct to determine $\langle|B_{||}|\rangle$ for only the unsaturated masers; i.e.,
 $\langle|B_{||}^{VLA1}|\rangle=(74\pm50)$~mG. The large errors are due to the large scatter of the magnetic field 
strength values, which depend on the projection effects, on the density of each water maser region, and on the local velocities of shocks. 
Because of this scatter it is difficult to evaluate the true strength of the full magnetic fields of the entire region. 
To compare the results obtained using different maser species we can estimate them roughly. 
Since $\langle\theta\rangle_{\rm{VLA1}}=83$\d$^{+7^{\circ}}_{-15^{\circ}}$ \, and $\langle\theta\rangle_{\rm{VLA2}}=85$\d$
^{+6^{\circ}}_{-36^{\circ}}$ the absolute weighted magnetic field 
strengths are  $|B_{\rm{VLA1}}|=\langle|B_{||}^{\rm{VLA1}}|\rangle/cos\langle\theta\rangle_{\rm{VLA1}}\sim700$\,mG and 
$|B_{\rm{VLA2}}|\sim1700$\,mG, respectively. Such high values are also found by Vlemmings et al. 
(\cite{vle06}) in Cepheus\,A. As their shock velocity is 10~\kms, measuring a magnetic field of 600~mG, these authors explain this high
 value with the presence of a nearby magnetic dynamo. \\
\indent If the magnetic field is important throughout the collapse of a spherical cloud, the cloud forms primarily by
flows along field lines, and the conservation of magnetic flux and of the mass imply $B\propto n^{0.47}$ (Crutcher \cite{cru99}). Although
the scaling of the magnetic field in water maser is due to shocks, Vlemmings (\cite{vle06}, \cite{vle08}) empirically showed that 
even water masers with number densities up to $10^{11} \rm{cm^{-3}}$ follow this relation, even if in non-masing gas of similar densities
 the magnetic field strengths are likely lower due to ambipolar diffusion. Hence,
from the magnetic field strengths, we can determine the number density of the cloud where the \hdueo \, masers arise by the 
relation $B\propto n^{0.47}$. Considering the pre-shock magnetic field 
($B_{||}=16$~mG) and density ($n_{\rm{H_{2}}}=10^{9}\,\rm{cm^{-3}}$) obtained by S09 from methanol maser emissions, we have 
$n_{\rm{H_{2}}}^{\rm{VLA1}}=3\times10^{10}\,\rm{cm^{-3}}$ and $n_{\rm{H_{2}}}^{\rm{VLA2}}=1\times10^{11}\,\rm{cm^{-3}}$. Similar 
calculation can be done considering the magnetic field strength obtained during the OH flare near VLA\,2 (Slysh et al. \cite{sly10}). 
They measured a magnetic field strength of about 70~mG, which is comparable to the total magnetic field strength inferred by S09 
($B=50$~mG). Since from the OH maser observations is generally possible to get only $B$ and not $B_{||}$ and since the flare 
occurred near VLA\,2, in this case we have to use the value of $|B_{\rm{VLA2}}|$ instead of the value of 
$\langle|B_{||}^{\rm{VLA2}}|\rangle$. For an OH number density of $10^{8}\,\rm{cm^{-3}}$, we find 
$n_{\rm{H_{2}}}^{\rm{VLA2}}\approx9\times10^{10}\,\rm{cm^{-3}}$ for the water masers around VLA\,2, close to what is obtained considering the 
methanol masers. These results,
which are close to the extreme upper limit of $10^{11}~\rm{cm^{-3}}$ of the water maser thermalisation (Elitzur et al. \cite{eli89}), 
confirm the assumptions on the \meth \, and OH number densities and reinforce the measurements of the magnetic field strength made by S09,
 Slysh et al. (\cite{sly10}) and in this paper. The three independent magnetic field strength values obtained from three different maser 
species located at different positions show that the magnetic field in massive star-forming regions can reach strength larger than 
expected. This large unexpected values of $B$ might be due either to the presence of a magnetic dynamo or to the presence of 
very high-velocity shocks that strongly compress the gas, and consequently increase the magnetic field strength from 50-70~mG (pre-shock 
methanol-OH region) to $>1$~Gauss (water maser region). For the last hypothesis we refer the reader to \S~\ref{shock}.\\
\indent With regard to the orientation of the magnetic field, it is worthwhile to discuss the influence of the Faraday 
rotation in our observations. 
\subsubsection{Faraday rotation}
\indent The measurements of the linear polarization angle ($\chi$) might be disturbed by the foreground, ambient, and internal Faraday 
rotation, which are given by
%S
\begin{equation}
\Phi[^{\circ}]=1.35\times10^{-15}\,D\,[\rm{cm}]\, \textit{n}_{e}\,[\rm{cm^{-3}}]\,  \textit{B}_{||}\,[\rm{mG}]\,\nu^{-2}\,[\rm{GHz}],
\label{fari}
\end{equation}
\noindent where $D$ is the length of the path over which the Faraday rotation occurs, $n_{\rm{e}}$ and $B_{||}$ are respectively the 
average electron density and the magnetic field along this path and $\nu$ the frequency. For the foreground Faraday rotation (i.e. 
the rotation due to the medium between the source and the observer) $\phi_{\rm{f}}$ is 0\d$\!\!.4$ at 22~GHz, assuming the interstellar 
electron density and the magnetic field are $n_{\rm{e}}\approx0.012\,\rm{cm^{-3}}$ and $B_{||}\approx2\,\rm{\mu G}$, respectively.\\
\indent All water masers are not in the same plane, but they arise in clouds at different depths. This depth is impossible to determine 
from our observations, so in order to estimate the ambient Faraday rotation (i.e. the rotation due to the ambient medium where masers 
arise), we assume the maximum projected separation as upper limit of $D$ among masers close to each other with the highest velocity 
difference, which is $D_{\rm{max}}\approx2\times10^{15}$\,cm. For the electron density, we assume the value reported by Fish \& Reid 
(\cite{fis06}), $n_{\rm{e}}\approx300\,\rm{cm^{-3}}$, which can produce a rotation of 90\d  \,along the path amplification of OH maser 
around H\,{\scriptsize II} regions. Considering $B_{||}=16$\,mG (S09), the ambient Faraday rotation is $\phi_{\rm{a}}<26$\d.\\
\indent The internal Faraday rotation, which can destroy the linear polarization (Fish \& Reid \cite{fis06}), depends on the type of 
shock that pumps the water masers. They can either be a dissociative jump shock (J-shock) or a nondissociative continuous shock 
(C-shock). In the case of a C-shock, the intrinsic Faraday rotation at 22~GHz can be considered negligible because the ionization state 
of the gas is controlled by cosmic-ray ionization, which generates electron-molecular ion pairs at a rate of $10^{-17}\rm{s^{-1}}$ 
(Kaufman \& Neufeld \cite{kau96}), while we have to roughly estimate its contribution for a J-shock. In this case, the electron 
density is $n_{\rm{e}}=\chi_{\rm{e}}\,n_{\rm{H_{2}}}$, where $\chi_{\rm{e}}$ is the ionization fraction ($\chi_{\rm{e}}=10^{-5}-10^{-4}$; 
Kylafis \& Norman \cite{kyl87}) and $n_{\rm{H_{2}}}$ the number density of the \hdueo \, masers. In the most conservative situation, for 
which $n_{\rm{e}}\sim10^{4}\,\rm{cm^{-3}}$, we determine an internal Faraday rotation of tens of degrees, enough to destroy the linear 
polarization in some cases. This means that either the electron density must be much lower or, more likely, the internal Faraday rotation 
must be negligible, and consequently the shock must be a C-shock.
\subsubsection {Magnetic field orientation}
Suppose that the water masers in both sources are pumped by C-shocks (see \S~\ref{shock}), this means that the Faraday rotation is 
$<26$\d. Since the \hdueo \, masers associated with VLA\,1 and VLA\,2 indicate that there are two different magnetic fields around the 
two sources, it is worthwhile discussing them separately. All water masers show $\theta>\theta_{\rm{crit}}\sim 55$\d; i.e., the magnetic 
field direction is perpendicular to the linear polarization vectors and close to the plane of the sky.\\
\indent The group VLA\,1A shows an orientation of the magnetic field $\varphi_{\rm{B}}^{\rm{VLA\,1A}}=70$\d$\pm9$\d, which is in good 
agreement with what is reported by S09 ($\varphi_{\rm{B}}=73$\d$\pm10$\d). This also confirms that the magnetic field orientation in this 
part of the source is close to the direction of the large-scale molecular bipolar outflow (66\d) (S09). The other two groups VLA\,1B and 
C show disordered linear polarization vectors, which cannot give information on the orientation of the magnetic field. The different 
orientation of the linear polarization vectors for these two groups might come from the different depths of the masers, i.e. high values 
of the ambient Faraday rotation, and therefore we can assume that the masers with angles far from $-20$\d \,are located at higher depths 
than those with angles close to $-20$\d \,(e.g. VLA1.14).\\
\indent Although around VLA\,2 some \hdueo ~masers show the 90\d-flip of the linear polarization angle (e.g., VLA2.15), which was also 
observed in some water masers in Cepheus\,A (Vlemmings et al. \cite{vle06}), the magnetic field appears to be radial
(see Fig.~\ref{vla2}). Considering $\langle \chi_{2}\rangle\approx-72$\d, the magnetic field orientation 
is about $18$\d \, that indicates a misalignment with the large-scale molecular bipolar outflow and the magnetic field of VLA\,1, even 
though the inclination of both ellipses ($\sim60$\d) is close to them. \\
\indent The different orientation of the magnetic fields of the two sources again implies different local physical environments for 
VLA\,1 and VLA\,2, and suggests that VLA\,1 is the powering source of the large-scale molecular bipolar outflow, as already indicated by S09.
\subsection{J- or C-shocks?}
\label{shock}
As reported in previous papers, the \hdueo \, masers are thought to be pumped by the transition of shocks generated in the two protostars 
(e.g., Torrelles et al. \cite{tor97}). These shocks can be either C-shocks or J-shocks. In a water masing region, C-shocks show 
$v\lesssim50$~\kms \, and a post-shock $T<4000$~K (Kaufman \& Neufeld \cite{kau96}), while J-shocks have velocities $v\gtrsim45$~\kms \, 
and a post-shock plateau temperature that depends on the pre-shock density, the shock velocity, and the observed linewidth of the water
 masers (Elitzur et al. \cite{eli89}). Kylafis \& Norman (\cite{kyl91}) investigated the effects of the temperature on their J-shock 
model up to a temperature of about 1000~K, and they found that the brightness temperature starts to increase more slowly when a temperature  
$\sim$500~K is reached. \\
\indent We first consider VLA\,1. Assuming the proper motion of the water masers reported by T03, i.e. 19~\kms, as the velocity of 
the shock and as pre-shock density, the density of the methanol masers $n_{\rm{H_{2}}}=10^{9}\rm{cm^{-3}}$ (S09), the equation reported 
in Elitzur et al. (\cite{eli89}) for a J-shock
\begin{equation}
T_{\rm{plateau}}\approx400\,\Bigr(\frac{n_{0}}{10^{7}\rm{cm^{-3}}}\,\frac{v_{\rm{s}}}{{100}\,\rm{km\,s^{-1}}}\frac{\Delta v^{-1}}{1\,\rm{km\,s^{-1}}}\Bigl)^{\frac{2}{9}}\,\rm{K},
\label{tpl}
\end{equation}
where here $\Delta v_{\rm{VLA1}}=0.5$~\kms, gives us a most likely post-shock temperature $T_{\rm{plateau}}^{\rm{VLA1}}\approx900$~K, 
which is higher than the temperature measured considering the weighted intrinsic velocity obtained from our fit 
($T_{\rm{VLA1}}\approx400$~K). Moreover, from the model of a C-shock described by Kaufman \& Neufeld (\cite{kau96}), we are able to 
estimate the velocity of the shock by considering the temperature $T_{\rm{VLA1}}$. The model for a temperature of 400~K gives 
$v_{\rm{s}}^{\rm{VLA1}}\sim15$~\kms, which is close to the proper motion of the \hdueo \, masers. Therefore, following our arguments, 
the shock in VLA\,1 is most likely a C-shock.\\
\indent In the case of VLA\,2, the plateau temperature due to a J-shock is $T_{\rm{plateau}}^{\rm{VLA2}}\approx1000\,$K 
(where $\Delta v_{\rm{VLA2}}=0.7$~\kms, and $v_{\rm{VLA2}}=46$~\kms), which is less than the temperature that we measured 
($T_{\rm{VLA2}}\approx2500$~K). The difference of the two temperatures may be due to the turbulence velocity of the gas, which can 
be determined by considering $\Delta V_{\rm{i}}$. In fact, $\Delta V_{\rm{i}}$ is related to the thermal and turbulence velocities by 
the equation
\begin{equation}
\Delta V_{\rm{i}}=\sqrt{\Delta V_{\rm{th}}^{2}+\Delta V_{\rm{turb}}^{2}},
\label{vel}
\end{equation}
and considering the Eq.~\ref{tv} we are able to estimate the turbulence velocity of the gas if a J-shock is present. The thermal velocity,
 which stems from the warming up of the gas by the compression due to the passage of a shock, for a temperature of 1000~K is 
$\Delta V_{\rm{th}}^{\rm{VLA\,2}}\approx1.6$~\kms, so we have $\Delta V_{\rm{turb}}\approx1.9$~\kms. If instead we 
suppose that the shock is a C-shock from the model of Kaufman \& Neufeld (\cite{kau96}) for $T_{\rm{VLA2}}\sim2500$~K, we get
 $v_{\rm{s}}^{\rm{VLA\,2}}\sim43$~\kms, which is close to the expansion velocity that we determined in the present paper 
($v_{\rm{VLA\,2}}=46$~\kms). Both velocities are also in good agreement with the velocity of $43$~\kms \, measured by Slysh et al. 
(\cite{sly10}) for the OH masers near VLA\,2 obtained by multiepoch observations. However, $v_{\rm{VLA\,2}}$ is close to the limit 
of 45~\kms, which is considered the limit between the two types of shocks, so we cannot rule one of them out by only considering the 
velocity. Since we obtain well-ordered high linear polarization fraction and linear polarization vectors, we have to suppose that the 
Faraday rotation is at most $\sim$20\d; otherwise, the vectors no longer appear to be aligned with the large structure 
(i.e., ellipse 2). This condition is only met if the internal Faraday rotation is small; in other words, if the shock is a C-shock 
(Kaufman \& Neufeld \cite{kau96}). Therefore, we strongly suggest that there is a C-shock also in VLA\,2 even if a J-shock cannot be 
completely ruled out.\\
\indent From the equation
\begin{equation}
\frac {B_{\rm{maser}}}{[\rm mG]}\approx80\sqrt{\frac{n_{\rm{H_{2}}}}{[10^{8}\,\rm{cm^{-3}}]}}\cdotp \frac{v_{\rm{s}}} {[10\,\rm{km\,s^{-1}}]},
\label{Cmag}
\end{equation}
where $n_{\rm{H_{2}}}$ is the pre-shock density and $v_{\rm{s}}$ the velocity of the shock, we can verify our assumption about the pre-shock 
density and the rough estimation of $|B_{\rm{VLA1}}|$ and $|B_{\rm{VLA2}}|$. Equation \ref{Cmag} is valid for both types 
of shocks (Kaufman \& Neufeld \cite{kau96}). Considering $n_{\rm{H_{2}}}=10^{9}\,\rm{cm^{-3}}$ as before, we
get $|B_{\rm{VLA1}}|\approx500$~mG and 
$|B_{\rm{VLA2}}|\approx1200$~mG, which are consistent with the rough values reported in \S~\ref{polma}, making the presence of a
magnetic dynamo unnecessary. Finally, from our results we can 
strongly depict the following scenario for VLA\,1: the hot gas in a C-shock pumps the water masers and at the same time the warm dust
 associated with this shock emits infrared photons, which pumps the methanol masers in the pre-shock region, as already suggested in S09.
\section{Conclusions}
We observed the 22~GHz water masers in full polarization mode in the massive star-forming region W75N with the VLBA. We detected 124 
water masers around the two radio sources VLA\,1 and VLA\,2, which appear to be in two different evolutionary stages; i.e.,
VLA\,1 is more evolved and a radio jet has not formed yet in VLA\,2. From our 
observations, we have shown that these two sources separated in the sky by $\sim1''$ (2000 AU) have different local physical environments.
 In particular, the linear polarization indicates that the magnetic field around VLA\,1 is aligned well with the large-scale molecular 
bipolar outflow, while the magnetic field is well-ordered around the shell-like expanding gas associated with VLA\,2. The detection of 
Zeeman-splitting indicates an absolute weighted magnetic field strengths $|B_{\rm{VLA1}}|\sim700$~mG and 
$|B_{\rm{VLA2}}|\sim1700$~mG. We were also able to determine the type of shocks that pump the water masers associated with both 
sources. They both are C-shocks, even though a J-shock cannot completely be ruled out in the case of VLA\,2. We suggest VLA\,1 as the 
driving source of the large-scale molecular bipolar outflow.\\

\noindent \small{\textit{Acknowledgments.} We wish to thank an anonymous referee and the editor for making useful suggestions that 
have highly improved the paper.
GS and WHTV acknowledge support by the Deutsche Forschungsgemeinschaft (DFG) through the Emmy 
Noether Reseach grant VL 61/3-1. SC acknowledges support from CONACYT grant 60581. JMT acknowledges support from MICINN (Spain)
 AYA2008-06189-C03 (co-funded with FEDER funds) and from Junta de Andaluc\'{\i}a (Spain).}

\end{document}